\documentclass[seceq]{ptptex}
\usepackage{wrapft}
\usepackage{graphicx}

\newcommand{\bra}[1]{\langle {#1} |}     
\newcommand{\ket}[1]{| {#1} \rangle}     
\newcommand{\bbra}[1]{\langle\!\langle {#1} |}     
\newcommand{\kket}[1]{| {#1} \rangle\!\rangle}     
\newcommand{\dket}[1]{|\!| {#1} \rangle}     
\newcommand{\wtilde}[1]{\widetilde{#1}} 

\newcommand{\ovl}[1]{\overline{#1}}
\newcommand{\lvect}[1]{\overleftarrow {#1}}
\newcommand{\rvect}[1]{\overrightarrow {#1}}

\def\beq{\begin{eqnarray}}
\def\eeq{\end{eqnarray}}
\def\bsub{\begin{subequations}}
\def\esub{\end{subequations}}
\def\b{\begin{equation}}

\title{
Scalar and Pseudoscalar Glueball Masses within\\
a Gaussian Wavefunctional Approximation
}

\author{
Yasuhiko {\sc Tsue}
}

\inst{
Physics Division, Faculty of Science, Kochi University, Kochi 780-8520, 
Japan
}

\recdate{
\today
}

\abst{
The lowest scalar and pseudoscalar glueball masses are evaluated by means of the time-dependent 
variational approach to the Yang-Mills gauge theory without fermions in the Hamiltonian 
formalism within a Gaussian wavefunctional approximation. 
The glueball mass is calculated as a pole of the propagator for a composite 
glueball field which consists of two massless gluons. 
The glueball propagator is here evaluated by using the linear response theory for 
the composite external glueball field. 
As a result, a finite glueball mass is obtained through the interaction between two massless gluons, 
in which the glueball mass depends on the 
QCD coupling constant $g$ in the nonperturbative form. 
}

\begin{document}

\maketitle

\section{Introduction}

The hadronic and/or quark-gluonic world governed by mainly the strong 
interaction reveals very fruitful physics such as existence of 
various possible phases, 
characteristic dynamical symmetries and various pattern of their 
symmetry breaking and so on in the hadronic and/or 
the quark-gluon matter.\cite{QGPreview}
In the hadronic world which 
should be basically described by the quantum chromodynamics (QCD), 
the color confinement is essential and the observed hadrons are mainly
ordinary ($3q$)-baryons and ($q{\bar q}$)-mesons. 
However, it is possible that the so-called color-singlet exotic hadrons 
exist such as ($3q(q{\bar q}))$-hadron like pentaquark hadron.\cite{nakano} 
It is believed that 
the color confinement occurs in QCD due to the non-Abelian nature\cite{asymptotic} 
of the 
gauge interaction.
This non-Abelian nature leads to the interaction between 
gauge fields themselves which represent gluons. 
Thus, it is interesting to consider a possible color-singlet state 
which only consists of gluons. 
This state is called the glueball.

The investigation about glueball has been performed widely in the 
theoretical side,\cite{review_th} 
for example, by using the bag model, the flux tube model, QCD sum rule method, 
the lattice QCD simulation and so on. 
Especially, some lattice QCD simulations have given 
the glueball masses with certain spins and parties. 
On the other hand, in the experimental side,\cite{review_ex} 
some candidates of glueball states have been reported. 
However, the glueball states mix with the normal ($q{\bar q}$)-meson states 
with the same spin and parity. 
Therefore, the definite 
information about glueballs such as masses and decay widths 
is not extracted experimentally until now.
Thus, the information about glueballs may be compared 
with the results obtained by the lattice QCD simulation.

The many theoretical investigations about glueball 
have been carried out by the use of the 
effective model of QCD. 
In their several investigation, the finite gluon mass is assumed in a 
certain treatment and the background of the model is not so clear 
because the gluon does not appear explicitly in a certain model. 
Thus, it may be necessary and interesting 
to deal with the glueball starting from the QCD Lagrangian itself.

In our previous paper\cite{0} which is refereed to as (I), 
the time-dependent variational method 
has been formulated for the Yang-Mills gauge theory without fermions in the 
Hamiltonian formalism, in which a Gaussian wavefunctional has been 
adopted as a possible trial state. 
This method presents an approximate treatment for the Yang-Mills 
gauge theory within the Gaussian wavefunctional approximation 
which corresponds to the Hartree approximation. 
Further, by the help of the linear response theory,\cite{DV} it may be possible 
that an approximation 
beyond the Hartree approximation, 
which may correspond to the random phase approximation (RPA) in 
many-body physics, is obtained in a non-perturbative way. 
In Ref.\citen{TVM00}, it was shown that the Goldstone theorem is satisfied in the 
time-dependent variational approach to the linear sigma model 
by the help of the linear response theory, while the Goldstone theorem breaks down in the 
tree-level approximation. 
Further, in Ref.\citen{TsueMatsuda}, the pion and sigma meson masses have been calculated 
by using the linear response theory in the linear sigma model. 
The same approach is possible to calculate the glueball masses 
in the QCD without quarks. 
Since the time-dependent variational method may be suitable 
for the use of the linear response theory in the quantum field theory, 
the Hamiltonian formalism is applied. 
In this paper, starting from the QCD Lagrangian density without quarks, 
the lowest scalar $(0^+)$ and pseudoscalar $(0^-)$ glueball masses 
are investigated. 
Then, the glueball masses are obtained reasonably compared with 
the results of the lattice QCD simulation. 
The glueball in this paper 
consists of two massless gluons which interact by the 
self-interaction due to the characteristic feature of QCD. 
As a result, through the interaction, the glueball gets mass.

This paper is organized as follows: 
In the next section, the time-dependent variational approach 
to the Yang-Mills gauge theory without fermions is summarized following (I). 
The formalism partially owes to Ref.\citen{Vautherin}. 
In \S 3, the method to calculate the glueball mass is explained. 
In \S 4, the lowest scalar and pseudoscalar glueball masses are 
evaluated in the modified minimal subtraction scheme which is 
described in detail in Appendix A. 
Also, the dependence of the glueball masses on the QCD coupling constant 
is shown. 
The last section is devoted to a summary and concluding remarks. 
In Appendix B, it is shown that the decay width may appear 
if an imaginary part of a response function investigated in this paper 
is considered seriously. 
In Appendix C, it is verified that 
the gluon mass is zero under the approximation used here in this formalism.

\section{Recapitulation of the time-dependent variational approach to 
the Yang-Mills gauge theory without fermions}

In this section, the time-dependent variational approach to the $su(N_c)$ 
Yang-Mills gauge theory, which 
has been developed in our paper,\cite{0} is given and summarized following 
(I) to make this paper be self-contained. 
In order to formulate the time-dependent variational method, the Hamiltonian formalism 
is adopted,\cite{KV} in which a Gaussian wavefunctional is applied as 
one of possible trial states.
This trial state includes a mean field and quantum fluctuations 
around it as variational 
functions. 
The time-development of the mean field and quantum fluctuations 
can be described under this Gaussian approximation.
The Gaussian state used here corresponds to a squeezed state.

\subsection{Hamiltonian formalism of Yang-Mills gauge theory without fermions}

As is well known, the $su(N_c)$ Yang-Mills gauge theory leads to a constrained system. 
Therefore, it is necessary to impose a constraint conditions. 
As was discussed in (I), as a result, the Hamiltonian density can be simply 
expressed\cite{1} 
so as to be in Eq.(I$\cdot$2$\cdot$10). 
Namely  
\beq\label{2-1}
{\cal H}_0=\frac{1}{2}\left[\left({\mib E}^a\right)^2+\left({\mib B}^a\right)^2\right]\ , 
\eeq
where roman letters $a$ denotes color indices. 
Here, we define ${\mib E}^a\equiv (E_x^a, E_y^a, E_z^a)$ and so on and 
\beq\label{2-2}
& &{\mib E}^a=-i\frac{\delta}{\delta {\mib A}^a}\ .
\nonumber\\
& &{\mib B}^a={\mib \nabla}\times {\mib A}^a-\frac{1}{2}gf^{abc}{\mib A}^b\times{\mib A}^c \ , 
\eeq
where ${\mib A}^a$ is a gauge field and its conjugate field is 
identical with a color-electric field 
${\mib E}^a$. 
Also, the color-magnetic field ${\mib B}^a$ is introduced and 
$f^{abc}$ is a structure constant of $su(N_c)$ Lie algebra.
In the functional Schr\"odinger representation\cite{JK} for a field theory,\cite{KV} 
the conjugate momentum field ${\mib E}^a$ in the gauge theory 
is represented as a functional derivative with respect to 
a gauge field ${\mib A}^a$, in which they obey canonical 
commutation relations: 
$[\ A_i^a({\mib x}),\ E_j^b({\mib y})\ ]=i\delta_{ij}\delta_{ab}\delta^3({\mib x}-{\mib y})$.

We formulate the time-dependent variational method for the Yang-Mils gauge theory. 
Then, it is necessary to introduce a trial state $\ket{\Phi}$ for variation. 
Here, in this paper, 
the trial wavefunctional is adopted as a Gaussian form as follows: 
\beq\label{2-3}
\Phi({\mib A}^a)&\equiv& \bra{{\mib A}^a}\Phi\rangle\nonumber\\
&=&{\cal N}^{-1}\exp\left(i\langle{\ovl {\mib E}}\ket{{\mib A}-{\ovl {\mib A}}}\right)
\exp\left(-\bra{{\mib A}-{\ovl {\mib A}}}\frac{1}{4G}-i\Sigma\ket{{\mib A}-{\ovl {\mib A}}}\right)
\ , 
\eeq
where ${\cal N}$ is a normalization factor and 
we use abbreviated notations as 
\beq\label{2-4}
& &\langle {\ovl {\mib E}}\ket{{\mib A}}=\int d^3{\mib x}
{\ovl {\mib E}}^a({\mib x},t)\cdot{\mib A}^a({\mib x}) \ , \nonumber\\
& &\bra{{\mib A}}\frac{1}{4G}\ket{{\mib A}}
=\int\!\int\! d^3{\mib x}\ d^3{\mib y}\ 
A^a_i({\mib x}) \frac{1}{4}G^{-1}{}^{ab}_{ij}
({\mib x},{\mib y},t)A^b_j({\mib y}) \ .
\eeq
Here, ${\ovl A}_i^a({\mib x},t)$, ${\ovl E}_i^a({\mib x},t)$, 
$G_{ij}^{ab}({\mib x},{\mib y},t)$ and 
$\Sigma_{ij}^{ab}({\mib x},{\mib y},t)$ are the variational functions 
in which ${\ovl A}_i^a({\mib x},t)$ and ${\ovl E}_i^a({\mib x},t)$ correspond 
to 
the expectation values of the field operators ${A}_i^a({\mib x})$ and 
$E_i^a({\mib x})$, respectively, 
and two-point functions $G_{ij}^{ab}({\mib x},{\mib y},t)$ and 
$\Sigma_{ij}^{ab}({\mib x},{\mib y},t)$ are related to the expectation values of the composite 
operators as follows: 
\beq\label{2-5}
& &\bra{\Phi}A_i^a({\mib x})\ket{\Phi}={\ovl A}_i^a({\mib x},t) \ , 
\nonumber\\
& &\bra{\Phi}E_i^a({\mib x})\ket{\Phi}={\ovl E}_i^a({\mib x},t) \ , 
\nonumber\\
& &\bra{\Phi}A_i^a({\mib x})A_j^b({\mib y})\ket{\Phi}=
{\ovl A}_i^a({\mib x},t){\ovl A}_j^b({\mib y},t)
+G_{ij}^{ab}({\mib x},{\mib y},t) \ , 
\nonumber\\
& &\bra{\Phi}E_i^a({\mib x})E_j^b({\mib y})\ket{\Phi}=
{\ovl E}_i^a({\mib x},t){\ovl E}_j^b({\mib y},t)
+\frac{1}{4}G^{-1}{}_{ij}^{ab}({\mib x},{\mib y},t)
+4(\Sigma G \Sigma)_{ij}^{ab}({\mib x},{\mib y},t) \ , \nonumber\\
& &\bra{\Phi}{A}_i^a({\mib x}){E}_j^b({\mib y})\ket{\Phi}=
{\ovl {A}}_i^a({\mib x},t){\ovl {E}}_j^b({\mib y},t)
+2(G \Sigma)_{ij}^{ab}({\mib x},{\mib y},t) \ . 
\eeq
Thus, it is understood that 
${\ovl A}_i^a({\mib x},t)$ represents a mean field for the field 
$A_i^a({\mib x})$
and a diagonal element of two-point function 
$G_{ij}^{ab}({\mib x},{\mib y},t)$, 
namely $G_{ii}^{aa}({\mib x},{\mib x},t)$, represents a quantum fluctuations 
around the mean field ${\ovl A}_i^a({\mib x},t)$.  
The state $\ket{\Phi}$ in (\ref{2-3}) is identical with the squeezed state.\cite{TF}
Thus, in this paper, the lowest excitation mode with a certain quantum number around vacuum is only treated.\cite{TsueMatsuda}

To determine the time-dependence of the state $\ket{\Phi}$ or 
a Gaussian wavefunctional $\Phi({\mib A}^a)$, it is necessary to 
determine the time-dependence of the variational functions 
${\ovl A}_i^a({\mib x},t)$, ${\ovl E}_i^a({\mib x},t)$, 
$G_{ij}^{ab}({\mib x},{\mib y},t)$ and 
$\Sigma_{ij}^{ab}({\mib x},{\mib y},t)$. 
The time-development of the state under the Hamiltonian density ${\cal H}$ 
is governed by the time-dependent variational principle 
in general: 
\beq\label{2-6}
\delta\int\! dt \bra{\Phi}i\frac{\partial}{\partial t}-\int\! d^3{\mib x}
{\cal H} \ket{\Phi}=0 \ .
\eeq
Here, the Hamiltonian derived from the Hamiltonian density (\ref{2-1}) 
can be expressed as 
\beq\label{2-7}
\langle H_0 \rangle&\equiv&
\int d^3{\mib x}\bra{\Phi}{\cal H}_0\ket{\Phi}\nonumber\\
&=&
\int d^3{\mib x}\biggl(
\frac{1}{2}{\ovl {\mib B}}^a({\mib x})\cdot {\ovl {\mib B}}^a({\mib x})
+\frac{1}{2}{\ovl {\mib E}}^a({\mib x})\cdot {\ovl {\mib E}}^a({\mib x})
+\frac{1}{8}{\rm Tr}\bra{\mib x}G^{-1}\ket{\mib x} \nonumber\\
& &\qquad\quad
+2{\rm Tr}\bra{\mib x}\Sigma G \Sigma \ket{\mib x}+
\frac{1}{2}{\rm Tr}\bra{\mib x}KG\ket{\mib x}
+\frac{g^2}{8}
\left({\rm Tr}[S_i T^a\bra{\mib x}G\ket{\mib x}]\right)^2 \nonumber\\
& &\qquad\quad
+\frac{g^2}{4}{\rm Tr}\left[S_iT^a\bra{\mib x}G\ket{\mib x}
S_iT^a\bra{\mib x}G\ket{\mib x}\right]\biggl) \ , 
\eeq
where
\beq\label{2-8}
& &{\ovl B}_i^a=\epsilon_{ijk}\partial_j {\ovl A}_k^a
-\frac{1}{2}gf^{abc}\epsilon_{ijk}{\ovl A}_j^b{\ovl A}_k^c \ , \nonumber\\
& &(S_i)_{jk}=i\epsilon_{ijk}\ , \qquad 
(T^a)^{bc}=-if^{abc} \ , \nonumber\\
& &K=(-i{\mib S}\cdot {\mib D})^2-g{\mib S}\cdot{\ovl {\mib B}} \ , \nonumber\\
& &{\mib D}=\nabla-ig{\ovl {\mib A}} \ , \qquad
{\ovl {A}}_i={\ovl A}_i^a T^a \ , \quad 
{\ovl {B}}_i={\ovl {B}}_i^a T^a \ .
\eeq
Here, $\epsilon_{ijk}$ is a complete antisymmetric tensor and 
${\mib S}$ implies a spin 1 matrix. 
The time-dependent variational principle (\ref{2-6}) leads to the 
following equations of motion: 
\bsub\label{2-9}
\beq
& &{\dot {\ovl {\mib A}}}^a({\mib x},t)
=\frac{\delta \langle H \rangle}{\delta {\ovl {\mib E}}^a({\mib x},t)}\ , 
\qquad
{\dot {\ovl {\mib E}}}^a({\mib x},t)
=-\frac{\delta \langle H \rangle}{\delta {\ovl {\mib A}}^a({\mib x},t)} \ , 
\label{2-9a}\\
& &{\dot G}_{ij}^{ab}({\mib x},{\mib y},t)
=\frac{\delta \langle H \rangle}{\delta \Sigma_{ij}^{ab}({\mib x},{\mib y},t)} 
\ , \qquad
{\dot \Sigma}_{ij}^{ab}({\mib x},{\mib y},t)
=-\frac{\delta \langle H \rangle}{\delta G_{ij}^{ab}({\mib x},{\mib y},t)} 
\ , 
\label{2-9b}
\eeq
\esub
where the dot represents a time-derivative, that is, 
${\dot {\ovl {\mib A}}}^a({\mib x},t)\equiv {\partial {\ovl {\mib A}}^a({\mib x},t) }/
\partial t$ and so on. 
These equations of motion are identical with the canonical equations of motion 
since the canonicity conditions,\cite{MMSK} which are developed in the theory of collective motion 
in nuclei, for the variational functions 
are implicitly imposed. 
Under the Hamiltonian density ${\cal H}_0$, a possible set of solutions 
with respect to 
${\ovl {\mib A}}^a$ and ${\ovl {\mib E}}^a$ is given by 
\beq\label{2-10}
{\ovl {\mib A}}^a({\mib x},t)={\ovl {\mib E}}^a({\mib x},t)={\mib 0} \ .
\eeq
%

Instead of the equations of motion for the two-point function 
$G_{ij}^{ab}({\mib x},{\mib y},t)$ and $\Sigma_{ij}^{ab}({\mib x},{\mib y},t)$ in 
Eq.(\ref{2-9b}), we can reformulate the equations of 
motion by introducing the reduced density matrix.\cite{TVM99} 
The reduced density matrix ${\cal M}$ is defined as 
\beq\label{2-11}
{\cal M}_{ij}^{ab}({\mib x},{\mib y},t)&=&
\left(\begin{array}{@{\,}cc@{\,}}
-i\langle {\hat A}_i^a({\mib x},t){\hat E}_j^b({\mib y},t)\rangle
-\frac{1}{2} & 
\langle {\hat A}_i^a({\mib x},t){\hat A}_j^b({\mib y},t)\rangle \\
\langle {\hat E}_i^a({\mib x},t){\hat E}_j^b({\mib y},t)\rangle & 
i\langle {\hat E}_i^a({\mib x},t){\hat A}_j^b({\mib y},t)\rangle-\frac{1}{2}
\end{array}\right) \nonumber\\
&=&
\left(\begin{array}{@{\,}cc@{\,}}
-2i(G\Sigma)_{ij}^{ab}({\mib x},{\mib y},t) & 
G_{ij}^{ab}({\mib x},{\mib y},t) \\
\frac{1}{4}(G^{-1})_{ij}^{ab}({\mib x},{\mib y},t)
+4(\Sigma G\Sigma)_{ij}^{ab}({\mib x},{\mib y},t) & 
2i(\Sigma G)_{ij}^{ab}({\mib x},{\mib y},t)
\end{array}\right) , \qquad\ \ 
\eeq
where ${\hat A}_i^a$ and ${\hat E}_i^a$ represent the quantum fluctuations around mean fields, 
which are defined by 
\beq\label{2-12}
{\hat A}_i^a({\mib x},t)&\equiv&A_i^{a}({\mib x})-\langle A_i^{a}({\mib x})
\rangle=A_i^a({\mib x})-{\ovl A}_i^a({\mib x},t) \ , \nonumber\\
{\hat E}_i^a({\mib x},t)&\equiv&E_i^{a}({\mib x})-\langle E_i^{a}({\mib x})
\rangle=E_i^a({\mib x})-{\ovl E}_i^a({\mib x},t) \ . 
\eeq
As for the reduced density matrix, it has been shown that 
the time-development of the reduced density matrix 
is governed by the following Liouville-von Neumann type equation of motion:
\bsub\label{2-13}
\beq
i{\dot {\cal M}}_{ij}^{ab}({\mib x},{\mib y},t)
&=&[\ {\wtilde {\cal H}}_0 \ , \ {\cal M}\ ]_{ij}^{ab}({\mib x},{\mib y},t) \ ,
\qquad\qquad\qquad\qquad\qquad\qquad
\label{2-13a}
\eeq
where the Hamiltonian matrix ${\wtilde {\cal H}}_0$ corresponding to ${\cal H}_0$ 
is introduced as 
\beq
{\wtilde {\cal H}}_0{}_{ij}^{ab}({\mib x},{\mib y},t)
&=&
\left(\begin{array}{@{\,}cc@{\,}}
0 & 
\delta_{ij}\delta_{ab} \\
\Gamma_{ij}^{ab}({\mib x},t) & 
0
\end{array}\right)\delta^3({\mib x}-{\mib y}) \ , 
\label{2-13b}\\
\Gamma_{ij}^{ab}({\mib x},t)&=&
K_{ij}^{ab}+g^2\left(
S_kT^c\bra{{\mib x}}G\ket{\mib x}S_kT^c\right)_{ij}^{ab} 
+\frac{g^2}{2}\left(S_kT^c\right)_{ij}^{ab}{\rm Tr}\left[
S_kT^c\bra{\mib x}G\ket{\mib x}\right]  \ .
\nonumber\\
& &
\label{2-13c}
\eeq
\esub

The reduced density matrix (\ref{2-11}) satisfies the following relation:
\beq\label{2-14}
{\cal M}^2=\left(\begin{array}{@{\,}cc@{\,}}
\frac{1}{4} & 0 \\
0 & \frac{1}{4}
\end{array}\right) \ . 
\eeq
Therefore, it is concluded that 
the eigenvalue of the reduced density matrix ${\cal M}$ itself is $+1/2$ or $-1/2$. 
Let the eigenstate for ${\cal M}$ be $\ket{\sigma 1/2,n,a,i}$ 
where $\sigma=\pm$ and $n$ represents a certain quantum number. 
Thus, the eigenvalue equation for ${\cal M}$ is written as 
\beq\label{2-15}
& &{\cal M}_{ij}^{ab}\ket{\sigma 1/2,n,b,j}=\sigma \frac{1}{2}\ket{\sigma 1/2,n,a,i} \ .  
\eeq
Then, let us introduce the mode functions $u$ and $v$ in the coordinate representation 
by using the eigenstates for ${\cal M}$:
\b\label{2-16}
\bra{\mib x}+1/2, n, a,i\rangle=\left(\begin{array}{@{\,}c@{\,}}
u_{n}{}_i^a({\mib x},t) \\
v_{n}{}_i^a({\mib x},t) 
\end{array}\right) \ .
\end{equation}
Then, it is possible to express the reduced density matrix 
in terms of the above mode 
functions, namely, the spectral decomposition can be carried out: 
\beq\label{2-17}
{\cal M}_{ij}^{ab}({\mib x},{\mib y},t)
&=&
\sum_{n (\sigma>0)} f_n
\biggl[ \left(\begin{array}{@{\,}c@{\,}}
u_{n}{}_i^a({\mib y},t) \\
v_{n}{}_i^a({\mib y},t) 
\end{array}\right)
(\ v_n^*{}_j^b({\mib y}) \ , \ u_n^*{}_j^b({\mib y})\ ) \nonumber\\
& &\qquad\qquad
+\left(\begin{array}{@{\,}c@{\,}}
u^*_{n}{}_i^a({\mib y},t) \\
-v^*_{n}{}_i^a({\mib y},t) 
\end{array}\right)
(\ -v_n{}_j^b({\mib y}) \ , \ u_n{}_j^b({\mib y})\ )
\biggl] \ ,
\eeq
where $f_n=1/2$. 
Here, the mode functions satisfy the following orthonormalized conditions as 
\beq\label{2-18}
& &\sum_{a,i}\int d^3{\mib x}\left(v^*_{n'}{}_i^a({\mib x})
u_{n}{}_i^a({\mib x})+u^*_{n'}{}_i^a({\mib x})v_{n}{}_i^a({\mib x})
\right)=\delta_{nn'} \ , 
\nonumber\\
& &\sum_{a,i}\int d^3{\mib x}\left(u^*_{n'}{}_i^a({\mib x})
v^*_{n}{}_i^a({\mib x})-v^*_{n'}{}_i^a({\mib x})u^*_{n}{}_i^a({\mib x})
\right)=0 \ , 
\nonumber\\
& &\sum_{a,i}\int d^3{\mib x}\left(u_{n'}{}_i^a({\mib x})
v_{n}{}_i^a({\mib x})-v_{n'}{}_i^a({\mib x})u_{n}{}_i^a({\mib x})
\right)=0 \ .
\eeq

For the equation of motion (\ref{2-13a}), if the reduced density matrix ${\cal M}={\cal M}_0$ 
has no time-dependence, the equation of motion is reduced to 
$[\ {\wtilde {\cal H}}_0\ , \ {\cal M}_0\ ]=0$, namely,
the reduced density matrix and the Hamiltonian matrix is commutable each other. 
In this situation, it is possible to diagonalize ${\cal M}_0$ and ${\wtilde {\cal H}}_0$ 
simultaneously. 
Thus, the diagonal basis $\ket{ai{\mib k}\sigma}$ 
can be introduced, where the following relations are satisfied:
\beq\label{2-19}
& &{\hat {\cal H}}_0\ket{ai{\mib k}\sigma}= E_{\bf k}^{\sigma}\ket{ai{\mib k}\sigma}
\ , \qquad
{\hat {\cal M}}_0\ket{ai{\mib k}\sigma}=f_{\bf k}^{\sigma}\ket{ai{\mib k}\sigma}
\ , \\
& &\qquad
E_{\bf k}^{\sigma}=\sigma E_{\bf k}\ , \qquad
f_{\bf k}^{\sigma}=\sigma\cdot\frac{1}{2}\ , \qquad
\sigma=\pm  \ . \nonumber
\eeq
Here, the second equation in Eq.(\ref{2-19}) is identical with Eq.(\ref{2-15}), 
so the diagonal basis is nothing but $\ket{\sigma 1/2,n={\mib k}, a, i}$:
\beq\label{2-20}
\ket{ai{\mib k}\sigma}=\ket{\sigma 1/2,{\mib k},a,i} \ .
\eeq
The Hamiltonian matrix and the reduced density matrix are diagonalized by 
using the 
mode functions in the momentum representation such as 
\beq\label{2-21}
U^{-1}{\cal H}_0 U=\left(\begin{array}{@{\,}cc@{\,}}
E_{\bf k} & 0 \\
0 & -E_{\bf k}
\end{array}\right)\ , \qquad
U^{-1}{\cal M}_0 U=\left(\begin{array}{@{\,}cc@{\,}}
\frac{1}{2} & 0 \\
0 & -\frac{1}{2}
\end{array}\right)\ . 
\eeq
Here, a unitary matrix $U$ in the momentum space is obtained as 
\beq\label{2-22}
U({\mib k})_{ij}^{ab}=
\left(\begin{array}{@{\,}cc@{\,}}
u({\mib k})_{ij}^{ab} & u^*({\mib k}){}_{ij}^{ab} \\
v({\mib k})_{ij}^{ab} & -v^*({\mib k}){}_{ij}^{ab}
\end{array}\right)\ , \qquad
U^{-1}({\mib k})_{ij}^{ab}=
\left(\begin{array}{@{\,}cc@{\,}}
v^*({\mib k}){}_{ij}^{ab} & u^*({\mib k}){}_{ij}^{ab} \\
v({\mib k})_{ij}^{ab} & -u({\mib k}){}_{ij}^{ab}
\end{array}\right)\ , \qquad
\eeq
where
\beq\label{2-23}
& &u({\mib k})_{ij}^{ab}=u_{{\bf k}}{}_i^a\delta_{ij}\delta_{ab}\ , \qquad
v({\mib k})_{ij}^{ab}=v_{{\bf k}}{}_i^a\delta_{ij}\delta_{ab}\ , 
\\
& &\left(\begin{array}{@{\,}c@{\,}}
u_{\bf k}{}_i^a \ , v_{\bf k}{}_i^a
\end{array}\right)
=\bra{\mib k}1/2, {\mib k}, a,i\rangle=\left(\frac{1}{\sqrt{2E_{\bf k}}} \ , 
\sqrt{\frac{E_{\bf k}}{2}} \ \right) \ ,
\nonumber\\
& &E_{\bf k}=|{\mib k}|\ . \nonumber
\eeq
In the diagonal basis, the Hamiltonian matrix and the reduced density matrix 
are expressed in the forms of the diagonal matrices as (\ref{2-21}). 
Originally, both are expressed in the form of Eqs.(\ref{2-13b}) and (\ref{2-11}), 
respectively. 
The relation between the original basis $\{ \dket{\alpha}\}$, 
which we implicitly used in Eqs.(\ref{2-13b}) and (\ref{2-11}), 
and the diagonal basis 
$\{ \ket{ai{\mib k}\sigma}\}$ 
is given by the relation 
\beq\label{2-24}
\dket{\alpha}\equiv U^{-1}\ket{ai{\mib k}\sigma} \ .
\eeq

\section{Scalar and pseudoscalar glueball masses}

In this section, a scalar and pseudoscalar glueball masses are calculated in the framework of the 
time-dependent variational method within the Gaussian approximation developed in the previous section. 
In general, the propagator $S$ can be obtained by using the generating function of the connected Green function, $W[J]$, as 
\beq\label{3-1}
S=\frac{\delta^2 W[J]}{\delta J\delta J}\ , 
\eeq
where $J$ represents a source current. Here, since the expectation value of the field operator $\varphi$ is 
obtained as $\varphi=\delta W[J] /\delta J$, the propagator can 
be expressed as 
\beq\label{3-2}
S_{IJ}=\frac{\delta \varphi_I}{\delta J_J}\ , 
\eeq
where the subscripts $I$ and $J$ imply certain indices. 
In our Hamiltonian formalism, we introduce the external field 
$\varphi_I({\mib x})$ with a source current $J_I(x)$, and the 
external Hamiltonian $H_{\rm ext}$ should be added to the Hamiltonian $H_0$: 
\beq\label{3-3}
H_{\rm ext}=\int d^3{\mib x}J_I({\mib x},t)\varphi_I({\mib x})
=\epsilon e^{-i\omega t}\int d^3{\mib x} e^{i{\bf q}\cdot{\bf x}}{J}_{I}{\varphi}_I({\mib x})\ , 
\eeq
where 
the source current $J_I(x)$ is assumed as $J_I(x)=\epsilon e^{-iqx}J_I$. 
Thus, as for 
the expectation value of the field operator $\varphi_I$ which we write 
$\langle \varphi_I\rangle$, the same time- and coordinate-dependence 
as that of the source current is expected: 
\beq\label{3-4}
\langle {\varphi}_I({\mib x}) \rangle =\beta_{I} e^{-i\omega t+i{\bf q}\cdot{\bf x}}\ .
\eeq
Therefore, the propagator $S_{IJ}$ is evaluated by  
\beq\label{3-5}
S_{IJ}=\lim_{\epsilon\rightarrow 0}\frac{\beta_{I}}{\epsilon J_J}\ . 
\eeq
Namely, a mass of particle represented by the field operator $\varphi_I$ is calculated as 
a pole of the propagator, namely, the pole of 
$\lim_{\epsilon\rightarrow 0}\beta_I/\epsilon$.\cite{TVM00}

Under the existence of the source current $J_I$, the reduced density matrix ${\cal M}$ is 
shifted from ${\cal M}_0$ which means the reduced density matrix without the source term. 
Of course, the Hamiltonian matrix ${\wtilde {\cal H}}$ is also shifted:  
\beq\label{3-6}
& &{\wtilde {\cal H}}={\wtilde {\cal H}}_0+\delta{\wtilde {\cal H}}\ , \nonumber\\
& &{\cal M}={\cal M}_0+\delta {\cal M}\ , 
\eeq
where the quantities with subscript 0 represent those without the source term. 
Since the change of the reduced density matrix leads to the change of its (1,2)-component, $G_{ij}^{ab}$ in Eq.(\ref{2-11}), 
then the Hamiltonian matrix which contains $G_{ij}^{ab}$ is also changed through $\Gamma_{ij}^{ab}$ in 
Eqs.(\ref{2-13b}) and (\ref{2-13c}), namely 
$\Gamma_{ij}^{ab}\rightarrow \Gamma_{ij}^{ab}+\delta\Gamma_{ij}^{ab}$. 
Thus, up to the order of 
$\epsilon$, the Liouville von-Neumann type equation of motion 
for the reduced density matrix can be expressed as  
\beq\label{3-7}
& &i\delta{\dot {\cal M}}=[\ {\wtilde {\cal H}}_0\ , \ \delta{\cal M}\ ]
+[\ \delta{\wtilde {\cal H}}_{\rm ext}\ , \ {\cal M}_0\ ]
+[\ \delta{\wtilde {\cal H}}_{\rm ind}\ , \ {\cal M}_0\ ]\ , \\
& &\delta{\wtilde {\cal H}}_{\rm ind}=
\left(\begin{array}{@{\,}cc@{\,}}
0 & \delta_{ij}\delta_{ab} \\
\delta\Gamma_{ij}^{ab} & 0
\end{array}\right)\ .
\nonumber
\eeq
Here, the shift of the Hamiltonian matrix, $\delta{\wtilde {\cal H}}$, is divided into two parts, 
namely $\delta {\wtilde {\cal H}}=\delta{\wtilde {\cal H}}_{\rm ind}+\delta{\wtilde {\cal H}}_{\rm ext}$, 
in which 
$\delta{\wtilde {\cal H}}_{\rm ind}$ represents the shift due to $\delta \Gamma$ and 
$\delta{\wtilde {\cal H}}_{\rm ext}$ represents the shift occurring 
directly from the introduction of the external Hamiltonian $H_{\rm ext}$.

\subsection{Scalar glueball mass: $0^{+}$}

The scalar glueball field is constructed as the following composite operator: 
\beq\label{3-8}
\varphi_I\equiv \frac{1}{2}F_{\mu\nu}^aF_a^{\mu\nu}={E}_i^a{E}_i^a+B_i^aB_i^a \ .
\eeq
As for the external field $B_i^aB_i^a$, the response $\delta {\cal M}$ has the same form as that for 
$E_i^aE_i^a$ in our Gaussian approximation. 
Thus, we adopt the external Hamiltonian simply as follows: 
\beq\label{3-9}
H_{\rm ext}=\int d^3{\mib x}\int d^3{\mib y}{J}_{ij}^{ab}({\mib x},{\mib y},t)E_i^a({\mib x})E_j^b({\mib y})\ , 
\eeq
where the source current $J_{ij}^{ab}({\mib x},{\mib y},t)$ is proportional to 
$\delta_{ij}\delta_{ab}\delta^3({\mib x}-{\mib y})$ for $0^+$ glueball state.  
We can derive the Liouville-von Neumann type equation of motion for the reduced density matrix 
as is similar to Eq.(\ref{2-13a}). 
\beq\label{3-10}
i{\dot {\cal M}}_{ij}^{ab}({\mib x},{\mib y},t)
&=&
[\ {\wtilde {\cal H}}_J\ , \ {\cal M}\ ]_{ij}^{ab}({\mib x},{\mib y},t)\ , \\
{\wtilde {\cal H}}_J&=&
{\wtilde {\cal H}}_0+\delta{\wtilde {\cal H}}_{\rm ext}+\delta{\wtilde {\cal H}}_{\rm ind}\ , \nonumber
\eeq
where 
\beq\label{3-11}
\delta{\wtilde {\cal H}}_{\rm ext}{}_{ij}^{ab}({\mib x},{\mib y},t)
=
\left(\begin{array}{@{\,}cc@{\,}}
0 & 2J_{ij}^{ab}({\mib x},{\mib y},t) \\
0 & 0
\end{array}\right) \ .  
\eeq

By using the equation $[\ {\wtilde {\cal H}}_0\ , \ {\cal M}_0\ ]=0$ and, first, omitting 
the induced term $\delta{\wtilde {\cal H}}_{\rm ind}$, the following equation of motion is obtained: 
\beq\label{3-12}
i\delta{\dot {\cal M}}=[\ {\wtilde {\cal H}}_0\ \ , \ \delta{\cal M}\ ]
+[\ \delta{\wtilde {\cal H}}_{\rm ext}\ , \ {\cal M}_0\ ]\ .
\eeq
We call $\delta{\cal M}$ derived by neglecting the induced term the bare response.

Let us take the source current in the form
\beq\label{3-13}
J_{ij}^{ab}({\mib x},{\mib y},t)
={\wtilde J}_{ij}^{ab}e^{-i\omega t+i{\bf q}\cdot{\bf x}}\delta^3({\mib x}-{\mib y})\ .
\eeq
Since $\delta {\wtilde {\cal H}}_{\rm ext}$ is proportional to $e^{-i\omega t+i{\mib q}\cdot{\mib x}}$, 
then, the bare response $\delta{\cal M}$ is also proportional to $e^{-i\omega t+i{\mib q}\cdot{\mib x}}$, 
which leads to $i\delta {\dot {\cal M}}=\omega \delta{\cal M}$. 
Thus, by using the diagonal basis $\{ \ket{ai{\mib k}\sigma}\}$ which obeys the eigenvalue equations in (\ref{2-19}), 
Eq.(\ref{3-12}) is recast into 
\beq\label{3-14}
\bra{ai{\mib k}\sigma}\delta{\cal M}\ket{bj{\mib k}'\sigma'}
=\frac{f_{{\bf k}'}^{{\sigma}'}-f_{{\bf k}}^{\sigma}}{\omega-(E_{\bf k}^{\sigma}-E_{{\bf k}'}^{{\sigma}'})}
\bra{ai{\mib k}\sigma}\delta{\wtilde {\cal H}}_{\rm ext}\ket{bj{\mib k}'\sigma'}\ .
\eeq
Here, we can rewrite the matrix elements in terms of the original basis as 
\beq\label{3-15}
\bra{ai{\mib k}\sigma}\delta{\wtilde {\cal H}}_{\rm ext}\ket{bj{\mib k}'\sigma'}
&=&
\sum_{\alpha,\beta}\bra{ai{\mib k}\sigma}\alpha\rangle\!\rangle \bbra{\alpha}
\delta{\wtilde {\cal H}}_{\rm ext}\kket{\beta}\langle\!\langle \beta\ket{bj{\mib k}'\sigma'}\nonumber\\
&=&U^{-1}({\mib k})_{il}^{ae}
\left(\begin{array}{@{\,}cc@{\,}}
0 & 2J_{lm}^{ef} \\
0 & 0
\end{array}\right)
U({\mib k}')_{mj}^{fb}\nonumber\\
&=&
\left(\begin{array}{@{\,}cc@{\,}}
2v({\bf k})^*{}_{il}^{ae}{\wtilde J}_{lm}^{ef}v({{\bf k}'}){}_{mj}^{fb} & 
-2v({\bf k})^*{}_{il}^{ae}{\wtilde J}_{lm}^{ef}v({{\bf k}'})^*{}_{mj}^{fb}
\\
2v({\bf k}){}_{il}^{ae}{\wtilde J}_{lm}^{ef}v({{\bf k}'}){}_{mj}^{fb} & 
-2v({\bf k}){}_{il}^{ae}{\wtilde J}_{lm}^{ef}v({{\bf k}'})^*{}_{mj}^{fb}
\end{array}\right)_{\!\!\sigma\sigma'}\!\!\!\!\bra{\mib k}e^{-iqx}\ket{{\mib k}'}
\nonumber\\
&=&\left[\left(\begin{array}{@{\,}c@{\,}}
v_{\bf k}^*{}_i^a \\
v_{\bf k}{}_i^a
\end{array}\right) 2{\wtilde J}_{ij}^{ab}(q)
\left(\begin{array}{@{\,}cc@{\,}}
v_{{\bf k}'}{}_j^b & -v_{{\bf k}'}^*{}_j^b
\end{array}\right)\right]_{\sigma\sigma'}\!\!
\delta^3({\mib k}'-{\mib k}+{\mib q})e^{-i\omega t} .
\eeq
As is similar to the above transformation, inversely, the shift of the reduced density matrix $\delta{\cal M}$ 
can be expressed in terms of the diagonal basis as 
\beq\label{3-16}
\delta{\cal M}&=&
\left(\begin{array}{@{\,}cc@{\,}}
-i\delta\langle {\hat A}_i^a{\hat E}_j^b\rangle
 & 
\delta\langle {\hat A}_i^a{\hat A}_j^b\rangle \\
\delta\langle {\hat E}_i^a{\hat E}_j^b\rangle & 
i\delta\langle {\hat E}_i^a{\hat A}_j^b\rangle
\end{array}\right) 
\nonumber\\
&=&U({\mib k})_{il}^{ae}
\left(\begin{array}{@{\,}cc@{\,}}
\bra{el{\mib k}+}\delta{\cal M}\ket{mf{\mib k}'+} & 
\bra{el{\mib k}+}\delta{\cal M}\ket{mf{\mib k}'-}  \\
\bra{el{\mib k}-}\delta{\cal M}\ket{mf{\mib k}'+}  & 
\bra{el{\mib k}-}\delta{\cal M}\ket{mf{\mib k}'-}
\end{array}\right) 
U^{-1}({\mib k}')_{mj}^{fb}\ . \quad
\eeq
Thus, we obtain the shift of $G_{ij}^{ab}$ and so on such as  
\beq\label{3-17}
\delta\langle {\hat A}_i^a{\hat A}_j^b\rangle
&=&
u_{\bf k}{}_i^a\bra{ai{\mib k}+}\delta{\cal M}\ket{bj{\mib k}'+}u_{{\bf k}'}^*{}_j^b
+u_{\bf k}^*{}_i^a\bra{ai{\mib k}-}\delta{\cal M}\ket{bj{\mib k}'+}u_{{\bf k}'}^*{}_j^b
\nonumber\\
& &-u_{\bf k}{}_i^a\bra{ai{\mib k}+}\delta{\cal M}\ket{bj{\mib k}'-}u_{{\bf k}'}{}_j^b
-u_{\bf k}^*{}_i^a\bra{ai{\mib k}-}\delta{\cal M}\ket{bj{\mib k}'-}u_{{\bf k}'}{}_j^b\ , 
\nonumber\\
&\equiv&
\delta{\cal G}_{+-}{}_{ij}^{ab}({\mib k},{\mib k}-{\mib q})\delta^3({\mib k}'-{\mib k}+{\mib q})e^{-i\omega t}
\nonumber\\
\delta\langle {\hat E}_i^a{\hat E}_j^b\rangle
&=&
v_{\bf k}{}_i^a\bra{ai{\mib k}+}\delta{\cal M}\ket{bj{\mib k}'+}v_{{\bf k}'}^*{}_j^b
-v_{\bf k}^*{}_i^a\bra{ai{\mib k}-}\delta{\cal M}\ket{bj{\mib k}'+}v_{{\bf k}'}^*{}_j^b
\nonumber\\
& &+v_{\bf k}{}_i^a\bra{ai{\mib k}+}\delta{\cal M}\ket{bj{\mib k}'-}v_{{\bf k}'}{}_j^b
-v_{\bf k}^*{}_i^a\bra{ai{\mib k}-}\delta{\cal M}\ket{bj{\mib k}'-}v_{{\bf k}'}{}_j^b
\nonumber\\
&\equiv&
\delta{\cal G}_{-+}{}_{ij}^{ab}({\mib k},{\mib k}-{\mib q})\delta^3({\mib k}'-{\mib k}+{\mib q})e^{-i\omega t}
\eeq
with (\ref{3-14}) and (\ref{3-15}).

Next, let us include the induced term $\delta {\wtilde {\cal H}}_{\rm ind}$. 
First, the shift of $G_{ij}^{ab}({\mib x},{\mib x})$ is calculated from Eq.(\ref{3-17}) as 
\beq\label{3-18}
\delta G_{ij}^{ab}({\mib x},{\mib x})
&=&
\bra{\mib x}\delta ({\hat A}_i^a{\hat A}_j^b )\ket{\mib x}
=\int d^3{\mib k}\int d^3{\mib k}' 
\bra{\mib x}{\mib k}\rangle \bra{\mib k}\delta ({\hat A}_i^a{\hat A}_j^b )\ket{{\mib k}'}
\bra{{\mib k}'}{\mib x}\rangle\nonumber\\
&=&\int d^3{\mib k}\int d^3{\mib k}' 
\frac{e^{{i({\bf k}-{\bf k}'})\cdot{\bf x}}}{(2\pi)^3}\bra{\mib k}\delta ({\hat A}_i^a{\hat A}_j^b )\ket{{\mib k}'}
\nonumber\\
&=&
e^{i{\bf q}\cdot{\bf x}}\int\frac{d^3{\mib k}}{(2\pi)^3}\delta {\cal G}_{+-}{}_{ij}^{ab}({\mib k}, {\mib k}-{\mib q})e^{-i\omega t}\nonumber\\
&\equiv&\alpha_{ij}^{ab}(q)e^{-iqx}\ .
\eeq
Similarly, 
\beq\label{3-19}
\delta S_{ij}^{ab}({\mib x},{\mib x})&\equiv&
\bra{\mib x}\delta ({\hat E}_i^a{\hat E}_j^b )\ket{\mib x}
=\int d^3{\mib k}\int d^3{\mib k}' 
\bra{\mib x}{\mib k}\rangle \bra{\mib k}\delta ({\hat E}_i^a{\hat E}_j^b )\ket{{\mib k}'}
\bra{{\mib k}'}{\mib x}\rangle\nonumber\\
&=&\int d^3{\mib k}\int d^3{\mib k}' 
\frac{e^{{i({\bf k}-{\bf k}'})\cdot{\bf x}}}{(2\pi)^3}\bra{\mib k}\delta ({\hat E}_i^a{\hat E}_j^b )\ket{{\mib k}'}
\nonumber\\
&=&
e^{i{\bf q}\cdot{\bf x}}\int\frac{d^3{\mib k}}{(2\pi)^3}\delta {\cal G}_{-+}{}_{ij}^{ab}({\mib k}, {\mib k}-{\mib q})e^{-i\omega t}\nonumber\\
&\equiv&\beta_{ij}^{ab}(q)e^{-iqx}\ . 
\eeq
Thus, the shift of $\Gamma_{ij}^{ab}$ in the Hamiltonian matrix, $\delta \Gamma_{ij}^{ab}$ in Eq.(\ref{3-7}), is 
given in the momentum representation as  
\beq\label{3-20}
\bra{\mib k}\delta \Gamma_{ij}^{ab}\ket{{\mib k}'}
&=&
\delta\Gamma_{ij}^{ab}(\omega,{\mib q})\delta^3({\mib k}'-{\mib k}+{\mib q})e^{-i\omega t}\nonumber\\
\delta\Gamma_{ij}^{ab}(\omega,{\mib q})&=&
\int\frac{d^3{\mib k}}{(2\pi)^3}\biggl[
g^2\left(S_kT^c\cdot \delta{\cal G}_{+-}({\mib k},{\mib k}-{\mib q})\cdot S_k T^c\right)_{ij}^{ab}\nonumber\\
& &\qquad\qquad\qquad 
+\frac{g^2}{2}(S_kT^c)_{ij}^{ab}{\rm Tr}\ 
\left[S_kT^c\cdot \delta {\cal G}_{+-}({\mib k},{\mib k}-{\mib q})\right]\biggl]\ .
\eeq
Here, from Eq.(\ref{3-18}), we obtain 
$\alpha_{ij}^{ab}(q)\equiv \int\frac{d^3{\bf k}}{(2\pi)^3}\delta{\cal G}_{+-}{}_{ij}^{ab}({\bf k},{\bf k}-{\bf q})$. 
Thus, $\delta \Gamma_{ij}^{ab}(q)$ can be expressed as follows:  
\beq\label{3-21}
\delta\Gamma_{ij}^{ab}(\omega,{\mib q})&=&
g^2(S_k)_{il}(T^c)^{ae}\alpha_{lm}^{ef}(q)(S_k)_{mj}(T^c)^{fb}\nonumber\\
& &+\frac{g^2}{2}(S_k)_{ij}(T^c)^{ab}\left[
(S_k)_{lm}(T^c)^{ef}\alpha_{ml}^{fe}(q)\right]\ .
\eeq
Further, the induced term of the Hamiltonian matrix is obtained in the diagonal basis as
\beq\label{3-22}
\bra{ai{\mib k}\sigma}\delta{\wtilde {\cal H}}_{\rm ind}\ket{bj{\mib k}'\sigma'}
&=&U^{-1}({\mib k})_{il}^{ae}
\left(\begin{array}{@{\,}cc@{\,}}
0 & 0 \\
\delta\Gamma_{lm}^{ef} & 0
\end{array}\right)
U({\mib k}')_{mj}^{fb}\nonumber\\
&=&
\left(\begin{array}{@{\,}cc@{\,}}
u({\bf k})^*{}_{il}^{ae}{\delta\Gamma}(q)_{lm}^{ef}u({{\bf k}'}){}_{mj}^{fb} & 
u({\bf k})^*{}_{il}^{ae}{\delta\Gamma}(q)_{lm}^{ef}u({{\bf k}'})^*{}_{mj}^{fb}
\\
-u({\bf k}){}_{il}^{ae}{\delta\Gamma}(q)_{lm}^{ef}u({\bf k}'){}_{mj}^{fb} & 
-u({\bf k}){}_{il}^{ae}{\delta\Gamma}(q)_{lm}^{ef}u({{\bf k}'})^*{}_{mj}^{fb}
\end{array}\right)_{\sigma\sigma'}
\nonumber\\
& &\qquad\qquad\qquad\qquad\qquad\qquad\qquad\qquad
\times
\delta^3({\mib k}'-{\mib k}+{\mib q})e^{-i\omega t}
\nonumber\\
&=&
\left[\left(\begin{array}{@{\,}c@{\,}}
u_{\bf k}^*{}_i^a \\
-u_{\bf k}{}_i^a
\end{array}\right) \delta\Gamma_{ij}^{ab}(q)
\left(\begin{array}{@{\,}cc@{\,}}
u_{{\bf k}'}{}_j^b & u_{{\bf k}'}^*{}_j^b
\end{array}\right)\right]_{\sigma\sigma'}\!\!
\delta^3({\mib k}'-{\mib k}+{\mib q})e^{-i\omega t}\ . \nonumber\\
& & 
\eeq
Thus, from the equation of motion (\ref{3-7}), we can obtain the response $\delta{\cal M}$ in the 
diagonal basis in the same way that the bare response was derived in Eq.(\ref{3-16}): 
\beq\label{3-23}
& &\bra{ai{\mib k}\sigma}\delta {\cal M}\ket{bj{\mib k}'\sigma'}\nonumber\\
&=&\frac{f_{{\bf k}'}^{{\sigma}'}-f_{\bf k}^{\sigma}}{\omega-(E_{\bf k}^{\sigma}-E_{{\bf k}'}^{\sigma'})}
\left[\bra{ai{\mib k}\sigma}\delta {\wtilde {\cal H}}_{\rm ind}\ket{bj{\mib k}'\sigma'}
+\bra{ai{\mib k}\sigma}\delta {\wtilde {\cal H}}_{\rm ext}\ket{bj{\mib k}'\sigma'}\right]
\nonumber\\
&=&\frac{f_{{\bf k}'}^{{\sigma}'}-f_{\bf k}^{\sigma}}{\omega-(E_{\bf k}^{\sigma}-E_{{\bf k}'}^{\sigma'})}
\nonumber\\
& &\times
\left[
\left(\begin{array}{@{\,}c@{\,}}
u_{\bf k}^*{}_i^a \\
-u_{\bf k}{}_i^a
\end{array}\right) \delta\Gamma_{ij}^{ab}(q)
\left(\begin{array}{@{\,}cc@{\,}}
u_{{\bf k}'}{}_j^b & u_{{\bf k}'}^*{}_j^b
\end{array}\right) 
+\left(\begin{array}{@{\,}c@{\,}}
v_{\bf k}^*{}_i^a \\
v_{\bf k}{}_i^a
\end{array}\right) 2{\wtilde J}_{ij}^{ab}(q)
\left(\begin{array}{@{\,}cc@{\,}}
v_{{\bf k}'}{}_j^b & -v_{{\bf k}'}^*{}_j^b
\end{array}\right) \right]_{\sigma\sigma'}
\nonumber\\
& &\qquad\qquad\qquad\qquad\qquad\qquad\qquad\qquad\qquad
\times
\delta^3({\mib k}'-{\mib k}+{\mib q})e^{-i\omega t}\ .
\eeq

Now, let us return to evaluating the scalar glueball mass. 
Since the field operator representing the scalar glueball is 
in Eq.(\ref{3-8}), 
the source current (\ref{3-13}) should be applied as follows:
\beq\label{3-24}
{\wtilde J}_{ij}^{ab}=\epsilon \delta_{ij}\delta_{ab} \ .
\eeq
In this case, the structure of 
the Lorentz and color indices of the response is the same one as 
${\wtilde J}_{ij}^{ab}$ such as  
$\alpha_{ij}^{ab}\propto \delta_{ij}\delta_{ab}$. 
Namely, 
\beq\label{3-25}
\delta G_{ij}^{ab}(\omega,{\mib q})=\alpha_{ij}^{ab}(q)=\alpha_i(q)\delta_{ij}
\delta^{ab}\ , \qquad
\beta_{ij}^{ab}(q)=\beta_i(q)\delta_{ij}\delta^{ab} \ .
\eeq
Then, the shift $\delta\Gamma_{ij}^{ab}$ in Eq.(\ref{3-21}) 
is simply written as 
\beq\label{3-26}
\delta\Gamma_{ij}^{ab}(\omega,{\mib q})
&=&g^2(S_k)_{il}(T^c)^{ae}\alpha_l(S_k)_{lj}(T^c)^{eb}\nonumber\\
&=&g^2c_3\left(\sum_l\alpha_l-\alpha_i\right)\delta_{ij}\delta_{ab}\ . 
\eeq
Here, we used $(S_k)_{ll}=0$, $(T^c)^{ee}=0$ and $\sum_c(T^c)^{ae}(T^c)^{eb}=c_3\delta_{ab}$ with 
$c_3=3$ for $su(3)$-algebra.
Further, it should be noticed that $f_+=1/2$ and $f_-=-1/2$, which leads to 
$\bra{ai{\mib k}+}\delta {\cal M}\ket{bj{\mib k}'+}
=\bra{ai{\mib k}-}\delta {\cal M}\ket{bj{\mib k}'-}=0$. 
Substituting (\ref{3-17}) into (\ref{3-18}) with $\bra{ai{\mib k}\sigma}
\delta{\cal M}\ket{bj{\mib k}'\sigma'}$ obtained in Eq.(\ref{3-23}), 
we can express $\alpha_i$ in terms of $\delta\Gamma_{ij}^{ab}$ and 
${\wtilde J}_{ij}^{ab}$ or $\epsilon$. 
However, $\delta\Gamma_{ij}^{ab}$ is also expressed by $\alpha_j$ in Eq.(\ref{3-26}). 
As a result, we obtain 
\beq\label{3-27}
\alpha_{i}(q)=g^2c_3\left(\sum_l\alpha_l(q)-\alpha_i(q)\right)
{\wtilde \Pi}_0(\omega,{\mib q})
-\epsilon {\wtilde \Sigma}_0(\omega,{\mib q}) \ , 
\eeq
where we defined 
\beq\label{3-28}
& &{\wtilde \Pi}_0(\omega,{\mib q})={\wtilde \Pi}_0^{(-)}(\omega,{\mib q})-{\wtilde \Pi}_0^{(+)}(\omega,{\mib q})\ , \nonumber\\
& &{\wtilde \Sigma}_0(\omega,{\mib q})={\wtilde \Sigma}_0^{(-)}(\omega,{\mib q})-{\wtilde \Sigma}_0^{(+)}(\omega,{\mib q})\ , \nonumber\\
& &{\wtilde \Pi}_0^{(\pm)}(\omega,{\mib q})
=\int\frac{d^3{\mib k}}{(2\pi)^3}\frac{1}{4E_{\bf k}E_{{\bf k}-{\bf q}}}\frac{1}{\omega\pm(E_{\bf k}+E_{{\bf k}-{\bf q}})}\ , \nonumber\\
& &{\wtilde \Sigma}_0^{(\pm)}(\omega,{\mib q})
=\int\frac{d^3{\mib k}}{(2\pi)^3}\frac{1}{2}\frac{1}{\omega\pm(E_{\bf k}+E_{{\bf k}-{\bf q}})}\ .
\eeq
Thus, we can solve Eq.(\ref{3-27}) with respect to $\alpha_i(q)$: 
\beq\label{3-29}
\alpha_i(q)=\frac{\epsilon{\wtilde \Sigma}_0(\omega,{\mib q})}{2g^2c_3{\wtilde \Pi}_0(\omega,{\mib q})-1}\ .
\eeq
In the same way, we can obtain $\beta_i(q)$ as 
\beq\label{3-30}
& &\beta_i(q)=\frac{\epsilon}{1-2g^2c_3{\wtilde \Pi}_0(\omega,{\mib q})}\left[
g^2c_3{\wtilde \Sigma}_0(\omega,{\mib q})^2+{\wtilde \Xi}_0(\omega,{\mib q})(1-2g^2c_3{\wtilde \Pi}_0(\omega,{\mib q}))\right] , \ \ \\
& &{\wtilde \Xi}_0(\omega,{\mib q})\equiv {\wtilde \Xi}_0^{(-)}(\omega,{\mib q})-{\wtilde \Xi}_0^{(+)}(\omega,{\mib q})\ , \nonumber\\
& &{\wtilde \Xi}_0^{(\pm)}(\omega,{\mib q})
=\int\frac{d^3{\mib k}}{(2\pi)^3}\frac{E_{\bf k}E_{{\bf k}-{\bf q}}}{2}\frac{1}{\omega\pm(E_{\bf k}+E_{{\bf k}-{\bf q}})}\ .
\nonumber
\eeq
Thus, the scalar glueball propagator $S_{ii}^{aa}(\omega,{\mib q})$ is obtained in the general manner 
given in (\ref{3-5}): 
\beq\label{3-31}
S_{ii}^{aa}(q)=\lim_{\epsilon\rightarrow 0}\frac{\beta_i(q)}{\epsilon}\ . 
\eeq
From the relation (\ref{3-31}) with (\ref{3-30}), the pole of propagator 
$S_{ii}^{aa}(q)$  
gives the scalar glueball mass $M_{0^+}$: 
\beq\label{3-32}
& &1-2g^2c_3{\wtilde \Pi}_0(M_{0^+},{\mib 0})=0 \ . 
\eeq

\subsection{Pseudoscalar glueball mass: $0^{-}$}

The field operator of the pseudoscalar glueball is given by 
\beq\label{3-33}
\varphi_I\equiv 
\frac{1}{2}F_{\mu\nu}^a{\wtilde F}_a^{\mu\nu}=\frac{1}{2}\left({E}_i^a{B}_i^a+B_i^aE_i^a\right)\ , 
\eeq
where ${\wtilde F}_a^{\mu\nu}=(1/2)\epsilon^{\mu\nu\rho\sigma}F_{\rho\sigma}^a$. 
Thus, let us consider the external Hamiltonian with external current $J_{ij}^{ab}$ as 
\beq\label{3-34}
H_{\rm ext}=\int d^3{\mib x}\int d^3{\mib y}{J}_{ij}^{ab}({\mib x},{\mib y},t)
\cdot\frac{1}{2}\left(E_i^a({\mib x})B_j^b({\mib y})+B_i^a({\mib x})E_j^b({\mib y})\right)\ .
\eeq
Here, the relation $J_{ij}^{ab}({\mib x},{\mib y},t)=J_{ji}^{ba}({\mib y},{\mib x},t)$ is satisfied. 
As is similar to the case of the scalar glueball, the external Hamiltonian matrix in the equation of 
motion for the reduced density matrix is obtained like Eq.(\ref{3-11}):
\beq\label{3-35}
\delta{\wtilde {\cal H}}_{\rm ext}{}_{ij}^{ab}({\mib x},{\mib y},t)
=
\left(\begin{array}{@{\,}cc@{\,}}
{\ovl J}^{\dagger}{}_{ij}^{ab}({\mib x},{\mib y},t) & 0 \\
0 & {\ovl J}_{ij}^{ab}({\mib x},{\mib y},t)
\end{array}\right) \ , 
\eeq
where we define 
\beq\label{3-36}
& &{\ovl J}_{ij}^{ab}({\mib x},{\mib y},t)\equiv
({\mib S})_{ii'}\cdot{\rvect {\mib \nabla}}^{\bf x}J_{i'j}^{ab}({\mib x},{\mib y},t)\ , \nonumber\\
& &{\ovl J}^{\dagger}{}_{ij}^{ab}({\mib x},{\mib y},t)\equiv
J_{ij'}^{ab}({\mib x},{\mib y},t){\lvect {\mib \nabla}}^{\bf y}\cdot({\mib S})_{j'j}\ . 
\eeq
Here, ${\rvect {\mib \nabla}}^{\bf x}$ (${\lvect {\mib \nabla}}^{\bf y}$) means the derivative 
with respect to ${\mib x}$ (${\mib y}$) for the quantity on the right-hand (left-hand) side. 
As the same way deriving the external Hamiltonian matrix 
in the diagonal basis in Eq.(\ref{3-15}), 
we obtain 
\beq\label{3-37}
& &\bra{ai{\mib k}\sigma}\delta{\wtilde {\cal H}}_{\rm ext}\ket{bj{\mib k}'\sigma'}
\nonumber\\
&=&\left[\left(\begin{array}{@{\,}c@{\,}}
v_{\bf k}^*{}_i^a \\
v_{\bf k}{}_i^a
\end{array}\right) {\cal J}^{\dagger}{}_{ij}^{ab}(q)
\left(\begin{array}{@{\,}cc@{\,}}
u_{{\bf k}'}{}_j^b & u_{{\bf k}'}^*{}_j^b
\end{array}\right)
+
\left(\begin{array}{@{\,}c@{\,}}
u_{\bf k}^*{}_i^a \\
-u_{\bf k}{}_i^a
\end{array}\right) {\cal J}_{ij}^{ab}(q)
\left(\begin{array}{@{\,}cc@{\,}}
v_{{\bf k}'}{}_j^b & -v_{{\bf k}'}^*{}_j^b
\end{array}\right)
\right]_{\sigma\sigma'}
\nonumber\\
& &\qquad\qquad\qquad\qquad\qquad\qquad\qquad\qquad\qquad\qquad
\times
\delta^3({\mib k}'-{\mib k}+{\mib q})e^{-i\omega t}\ , 
\eeq
where
\beq\label{3-38}
& &{\cal J}_{ij}^{ab}(q)=i({\mib S})_{ii'}\cdot{\mib q}{\wtilde J}_{i'j}^{ab}\ , \qquad
{\cal J}^{\dagger}{}_{ij}^{ab}=-i{\wtilde J}_{ij'}^{ab}({\mib S})_{j'j}\cdot{\mib q}\ , \\
& &J_{ij}^{ab}({\mib x},{\mib y},t)
=\int d^3{\mib q}{\wtilde J}_{ij}^{ab}(q)e^{-i\omega t}e^{i{\bf q}\cdot{\bf x}}\delta^3({\mib x}-{\mib y})\ . 
\nonumber
\eeq
Including the induced term $\delta{\wtilde {\cal H}}_{\rm ind}$, the shift of the reduced density matrix 
is calculated in the same manner deriving Eq.(\ref{3-22}). 
As a result, we obtain 
\beq\label{3-39}
& &\bra{ai{\mib k}\sigma}\delta {\cal M}\ket{bj{\mib k}'\sigma'}\nonumber\\
&=&\frac{f_{{\bf k}'}^{{\sigma}'}-f_{\bf k}^{\sigma}}{\omega-(E_{\bf k}^{\sigma}-E_{{\bf k}'}^{\sigma'})}
\biggl[
\left(\begin{array}{@{\,}c@{\,}}
u_{\bf k}^*{}_i^a \\
-u_{\bf k}{}_i^a
\end{array}\right) \delta\Gamma_{ij}^{ab}(q)
\left(\begin{array}{@{\,}cc@{\,}}
u_{{\bf k}'}{}_j^b & u_{{\bf k}'}^*{}_j^b
\end{array}\right) 
\nonumber\\
& &\qquad\qquad
+
\left(\begin{array}{@{\,}c@{\,}}
v_{\bf k}^*{}_i^a \\
v_{\bf k}{}_i^a
\end{array}\right) {\cal J}^{\dagger}{}_{ij}^{ab}(q)
\left(\begin{array}{@{\,}cc@{\,}}
u_{{\bf k}'}{}_j^b & u_{{\bf k}'}^*{}_j^b
\end{array}\right)
+
\left(\begin{array}{@{\,}c@{\,}}
u_{\bf k}^*{}_i^a \\
-u_{\bf k}{}_i^a
\end{array}\right) {\cal J}_{ij}^{ab}(q)
\left(\begin{array}{@{\,}cc@{\,}}
v_{{\bf k}'}{}_j^b & -v_{{\bf k}'}^*{}_j^b
\end{array}\right)\biggl]_{\sigma\sigma'}
\nonumber\\
& &\qquad\qquad\qquad\qquad\qquad\qquad\qquad\qquad\qquad\qquad\qquad
\times\delta^3({\mib k}'-{\mib k}+{\mib q})e^{-i\omega t}\ .
\eeq

As for the pseudoscalar glueball, let us take ${\wtilde J}_{ij}^{ab}$ 
in the form 
\beq\label{3-40}
{\wtilde J}_{ij}^{ab}=\epsilon \delta_{ij}\delta_{ab}\ .
\eeq
Under this form, the source current appearing in Eq.(\ref{3-37}) has the following form: 
\beq\label{3-41}
& &{\cal J}_{ij}^{ab}(q)=i\epsilon ({\mib S})_{ij}\cdot {\mib q}\delta_{ab}=-\epsilon\epsilon_{kij}q_k\delta_{ab}\ , \\
& &{\cal J}^{\dagger}{}_{ij}^{ab}(q)=-{\cal J}_{ij}^{ab}(q)\ . \nonumber
\eeq
The response $\alpha_{ij}^{ab}(q)$in Eq.(\ref{3-18}) has the same 
structure with respect to the Lorentz and color indices as that 
of ${\cal J}_{ij}^{ab}(q)$, 
namely
\beq\label{3-42}  
\alpha_{ij}^{ab}(q)={\mib \alpha}\cdot({\mib S})_{ij}\delta_{ab}
= \alpha_k(q)\epsilon_{ijk}\delta_{ab} \ ,
\eeq
where $\alpha_k(q)$ is introduced. 
Following the same procedure in which $\alpha_i(q)$ in Eq.(\ref{3-27}) was derived, 
we can obtain the following relation as
\beq\label{3-43}
& &\alpha_{ij}^{ab}(q)=\delta\Gamma_{ij}^{ab}(q){\wtilde \Pi}_0(q)+\left(
{\cal J}^{\dagger}{}_{ij}^{ab}(q)-{\cal J}_{ij}^{ab}(q)\right){\wtilde \Upsilon}_0(q) \ , 
\eeq
where 
\beq\label{3-44}
\delta\Gamma_{ij}^{ab}(q)=g^2c_3\alpha_{ij}^{ab}(q)
\eeq
and we defined 
\beq\label{3-45}
& &{\wtilde \Upsilon}_0(q)={\wtilde \Upsilon}_0^{(-)}(q)+{\wtilde \Upsilon}_0^{(+)}(q)\ , \nonumber\\
& &{\wtilde \Upsilon}_0^{(\pm)}(q)
=\int\frac{d^3{\mib k}}{(2\pi)^3}\frac{1}{4E_{\bf k}}\frac{1}{\omega\pm(E_{\bf k}+E_{{\bf k}-{\bf q}})}\ .
\eeq
Thus, $\alpha_i(q)$ which was introduced in Eq.(\ref{3-42}) is determined: 
\beq\label{3-46}
\alpha_k=\frac{2\epsilon q_k{\wtilde \Upsilon}_0(q)}{1-g^2c_3{\wtilde \Pi}_0(q)}\ . 
\eeq
Instead of (\ref{3-19}), we should investigate the following quantity:
\beq\label{3-47}
\delta S_{ij}^{ab}&\equiv&
\bra{\mib x}\delta({\hat E}_i^a{\hat B}_j^b)\ket{\mib x}
=\epsilon_{jlm}q_le^{i{\bf q}\cdot{\bf x}}\int\frac{d^3{\mib k}}{(2\pi)^3}\delta{\cal G}_{--}{}_{im}^{ab}({\mib k},{\mib k}-{\mib q})
e^{-i\omega t}\nonumber\\
&\equiv&\epsilon_{jlm}q_l e^{-iqx}\beta_{im}^{ab}(q)\ , 
\eeq
where 
\beq\label{3-48}
\beta_{im}^{ab}(q)&\equiv&
\int\frac{d^3{\mib k}}{(2\pi)^3}\Bigl[v_{\bf k}{}_i^a\bra{ai{\mib k}+}\delta{\cal M}\ket{bm{\mib k}-{\mib q}+}u_{{\mib k}-{\mib q}}^*{}_m^b
\nonumber\\
& &\qquad
-v_{\bf k}^*{}_i^a\bra{ai{\mib k}-}\delta{\cal M}\ket{bm{\mib k}-{\mib q}+}u_{{\mib k}-{\mib q}}^*{}_m^b
-v_{\bf k}{}_i^a\bra{ai{\mib k}+}\delta{\cal M}\ket{bm{\mib k}-{\mib q}-}u_{{\mib k}-{\mib q}}{}_m^b\nonumber\\
& &\qquad
+v_{\bf k}^*{}_i^a\bra{ai{\mib k}-}\delta{\cal M}\ket{bm{\mib k}-{\mib q}-}u_{{\mib k}-{\mib q}}{}_m^b\Bigl]\nonumber\\
&=&
\delta\Gamma_{im}^{ab}(q){\wtilde \Upsilon}_0(q)
+{\cal J}^{\dagger}{}_{im}^{ab}{\wtilde \Psi}_0(q)-{\cal J}_{im}^{ab}{\wtilde \Psi}'_0(q)\ . 
\eeq
Here, we define 
\beq\label{3-49}
& &{\wtilde \Psi}_0(q)=\int\frac{d^3{\mib k}}{(2\pi)^3}\frac{E_{\bf k}}{4E_{{\bf k}-{\bf q}}}
\left(
\frac{1}{\omega-(E_{\bf k}+E_{{\bf k}-{\bf q}})}-\frac{1}{\omega+(E_{\bf k}+E_{{\bf k}-{\bf q}})}\right)\ , 
\nonumber\\
& &{\wtilde \Psi}'_0(q)=\int\frac{d^3{\mib k}}{(2\pi)^3}\frac{1}{4}
\left(
\frac{1}{\omega-(E_{\bf k}+E_{{\bf k}-{\bf q}})}-\frac{1}{\omega+(E_{\bf k}+E_{{\bf k}-{\bf q}})}\right)\ . 
\eeq
Finally, from Eqs.(\ref{3-44}) and (\ref{3-46}), we obtain the following response:
\beq\label{3-50}
\beta_{im}^{ab}(q)=\epsilon\epsilon_{imk}q_k\cdot 
\frac{2g^2c_3{\wtilde \Upsilon}_0^2(q)+
(1-g^2c_3{\wtilde \Pi}_0(q))({\wtilde \Psi}_0(q)+{\wtilde \Psi}'_0)}{1-g^2c_3{\wtilde \Pi}_0(q)}\delta_{ab} \ .
\eeq
The pseudoscalar glueball propagator is given as $\beta_{im}^{ab}/\epsilon$, so 
the pseudoscalar glueball mass $M_{0^-}$ is evaluated from the following: 
\beq\label{3-51}
& &1-g^2c_3{\wtilde \Pi}_0(M_{0^-},{\mib 0})=0 \ . 
\eeq
From Eq.(\ref{3-32}) and (\ref{3-51}), our final task is to calculate ${\wtilde \Pi}_0(q)$, which we call the 
polarization tensor, in order to get the glueball masses.

\section{Dependence of glueball masses on the QCD coupling constant}

The polarization tensor is rewritten in the form with the 4-momentum integration:
\beq\label{4-1}
g^2{\wtilde \Pi}_0(\omega,{\mib q})&\equiv& \Pi_0(\omega,{\mib q})\nonumber\\
&=&g^2\int\frac{d^3{\mib k}}{(2\pi)^3}\frac{1}{4E_{\bf k}E_{{\bf k}-{\bf q}}}
\left[\frac{1}{\omega-(E_{\bf k}+E_{{\bf k}-{\bf q}})}-\frac{1}{\omega+(E_{\bf k}+E_{{\bf k}-{\bf q}})}\right]\nonumber\\
&=&g^2\int\frac{d^4 k}{i(2\pi)^4}S_kS_{k-q}\ , \\
S_k&=&\frac{-i}{k^2+i\epsilon}\ . \nonumber
\eeq
The integral diverges. 
Thus, it is necessary to regularize the divergent integral to get a finite result. 
We here apply the dimensional regularization method. 
First, we introduce a momentum scale $\mu$ and define the dimensionless coupling $g_R$ instead of $g$ because the space-time dimension is now $n$. 
The detail calculation is given in Appendix A. 
As a result, we finally obtain the finite result for the polarization tensor: 
\beq\label{4-2} 
\Pi_0(q)\equiv g^2{\wtilde \Pi}_0(q)=\frac{g_R^2}{16\pi^2}\left(\ln\frac{q^2}{e^2\mu^2}-i\pi\right)\ .
\eeq
The imaginary part of the polarization tensor leads to the decay of glueball, namely, it gives the decay width of glueball. 
However, in this treatment in this paper, only gluon is contained in the theory. 
Then, it is only possible that the glueball can decay to the color-octet gluon, which should be forbidden due to 
the color confinement. 
Thus, we here neglect the imaginary part of the polarization tensor by hand 
because the color confinement is not fully taken into account in this treatment. 
It is shown in Appendix B that the imaginary part of the polarization tensor gives a rather large 
decay width under the same parameter set. 
Thus, under omitting the imaginary part, the scalar glueball mass $M_{0^+}$ and the pseudoscalar glueball mass $M_{0^-}$ are 
determined by the following relations given in Eqs.(\ref{3-32}) and (\ref{3-51}), respectively, as 
\beq\label{4-3}
& &1-a_{\pm}c_3{\Pi}_0(M_{0^{\pm}},{\mib 0})=0 \ ,  
\eeq
where $a_{+}=2$ for $0^+$ and $a_{-}=1$ for $0^-$ glueball. 
From Eq.(\ref{4-2}) under ignoring the imaginary part, 
the glueball masses are derived as a function of the QCD coupling constant $g_R$ as  
\beq
& &M_{0^+}=\mu e\cdot e^{\frac{4\pi^2}{3g_R^2}}
=\Lambda_{\rm QCD}\cdot\exp\left(1+\frac{4\pi^2}{g_R(\mu)^2}\left(\frac{1}{3}+\frac{1}{4\pi^2b_0}\right)\right) \ ,
\label{4-5}\\
& &M_{0^-}=\mu e\cdot e^{\frac{8\pi^2}{3g_R^2}}
=\Lambda_{\rm QCD}\cdot\exp\left(1+\frac{4\pi^2}{g_R(\mu)^2}\left(\frac{2}{3}+\frac{1}{4\pi^2b_0}\right)\right)\ , 
\label{4-6}
\eeq
where the renormalization-group-invariant QCD scale parameter $\Lambda_{\rm QCD}$ is introduced. 
Although it may not be necessary to regard $g_R$ as a running coupling in the nonperturbative variational treatment, 
it is known that the running coupling can be derived 
in the lowest order approximation in this variational formalism.\cite{Vautherin} 
Thus, let the QCD coupling constant $g_R(\mu)$ be regarded as a running coupling constant depending on the momentum scale $\mu$:
\beq\label{4-7}
\Lambda_{\rm QCD}=\mu e^{-\frac{1}{b_0g_R^2(\mu)}}\ , \qquad
g_R^2(\mu)=\frac{1}{\frac{b_0}{2}\ln\frac{\mu^2}{\Lambda_{\rm QCD}^2}}\ , \qquad
b_0\equiv \frac{1}{8\pi^2}\cdot\frac{11N_c-2N_f}{3}\ . \ \ 
\eeq
Here, $N_c$ and $N_f$ are the number of color and flavor, respectively. 
We here adopt $N_c=3$ and $N_f=0$ for the pure gauge theory, which leads to 
$b_0=11/8\pi^2$.

\begin{figure}[t]
\begin{center}
\includegraphics[height=5.4cm]{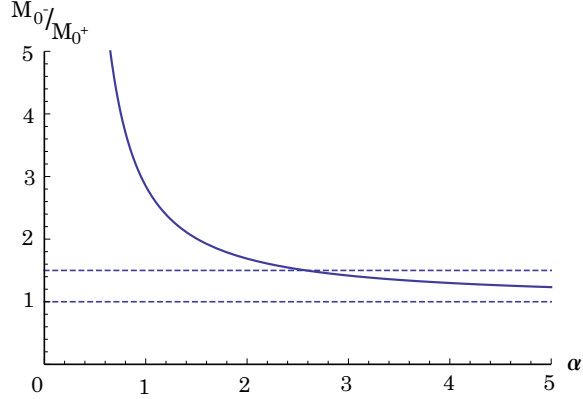}
\caption{
The mass ratio between the scalar $(M_{0^+})$ and pseudoscalar ($M_{0^-})$ glueball 
masses is shown as a function of the QCD running coupling 
$\alpha_{\rm QCD}=g_R(\mu)^2/(4\pi)$. The dotted lines represent the ratio 1 and 1.5, respectively. 
}
\label{fig:4-1}
\end{center}
\end{figure}

\begin{figure}[t]
\begin{center}
\includegraphics[height=5.1cm]{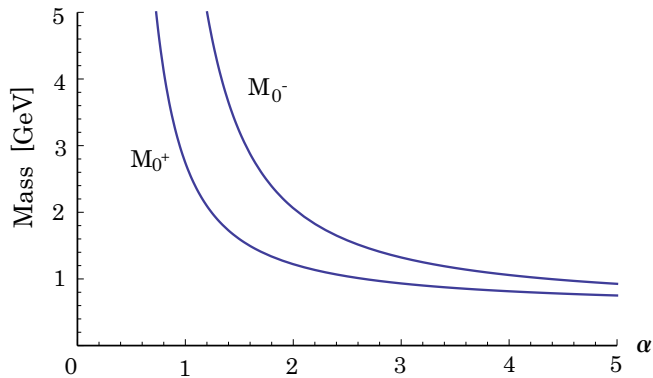}
\caption{
The scalar glueball mass $(M_{0^+})$ and the pseudoscalar glueball mass ($M_{0^-})$ 
are shown as a function of the QCD running coupling 
$\alpha_{\rm QCD}=g_R(\mu)^2/(4\pi)$. 
The QCD scale parameter is taken as $\Lambda_{\rm QCD}=0.20$ GeV.
}
\label{fig:4-2}
\end{center}
\end{figure}

In Fig.1, the mass ratio is shown as a function of the QCD coupling $\alpha_{\rm QCD}=g_R(\mu)^2/(4\pi)$. 
The glueball masses themselves are shown in Fig.2 as a function of $\alpha_{\rm QCD}$. 
Here, the QCD scale parameter is adopted as $\Lambda_{\rm QCD}=0.20$ GeV. 
For example, if the QCD coupling $\alpha_{\rm QCD}$ is roughly taken as $\alpha_{\rm QCD}=1.6$, then 
the glueball masses are obtained as 
\beq\label{4-8}
M_{0^+}=1.50\ {\rm GeV}\ , \qquad
M_{0^-}=2.88\ {\rm GeV}\ .
\eeq
We cannot compare these values with experimental meson masses directly because 
the glueball must mix the scalar or pseudoscalar $q{\bar q}$-mesons, 
while there exist glueball candidates in particle data. 
So, the glueball masses obtained in our framework should be compared with the lattice QCD calculation 
such as $M_{0^+}=1.71$ GeV and $M_{0^-}=2.56$ GeV.\cite{lattice} 
Further, if we take $\Lambda_{\rm QCD}=0.25$ GeV, then, $M_{0^+}=1.53$ GeV and $M_{0^-}=2.58$ GeV 
for $\alpha_{\rm QCD}=2.0$ are obtained.
Roughly speaking, the result in this paper is not so bad.

\section{Summary and concluding remarks}

The scalar and the pseudoscalar glueball masses have been investigated in the framework 
of the time-dependent variational method and the linear response theory. 
We have started with the Hamiltonian of the $su(3)$ Yang-Mills gauge theory without fermions, 
namely QCD Hamiltonian without quarks. 
The time-dependent variational method within the Gaussian wavefunctional, 
which includes the mean field and quantum fluctuations around it, has been formulated 
in order to evaluate the glueball mass. 
The glueball mass has been calculated as the pole mass of the propagator of glueball. 
Here, the glueball propagator has been derived from the response with respect to the external 
composite field representing the glueball, 
which consists of two massless gluons. 

The gluon mass itself is zero in this method. 
Thus, it is shown that 
the finite glueball mass is properly generated through the interaction between massless gluons. 
Further, since the dependence of the glueball masses 
on the QCD coupling constant $g$ reveals the form 
$1/g^2$, the results may not be arrived by the perturbation theory 
with respect to $g$.

In this paper, the coupling dependence of glueball masses was given. 
In the renormalization group calculation, the glueball masses $m_G$ are expressed as 
\beq\label{5-1}
m_G=c_G\mu\exp\left(\int\frac{dg}{\beta(g)}\right)=c_G\Lambda_{\rm QCD}\ , 
\eeq
where $c_G$ is a constant.\cite{RG}
Further, the similar expression was obtained in the context of the asymptotic limit in the lattice gauge theory.\cite{Strong1}
In the strong coupling expansion of the lattice QCD, it was obtained that the glueball masses decrease when the coupling $g$ decreases,\cite{Strong1} 
while the change of mass ratio between $0^+$ state and $1^+$ or $2^+$ state, instead of $0^{-}$ state, is similar to our result, 
namely the mass ratio increases when $g$ decreases.\cite{Strong2}  
Further, in the large $N_c$ limit in the gauge/string duality, it seems that there is a tendency that the glueball masses decrease 
as $g$ decreases.\cite{string}
These results seem to be different from our result in which the glueball masses increase when $g$ decreases. 
However, it may be natural that, in our result, the glueball masses become very large in the asymptotic region with very small $g$, namely in the quark-gluon phase, 
because it may be impossible that the glueballs are excited and produced in the deconfined phase.  
The investigation of the implication to the strong coupling expansion or gauge/string duality is an interesting future problem. 

Experimentally, it is difficult to extract the glueball masses properly because the glueball states 
mix the other $(q{\bar q})$-meson states with same quantum numbers. 
Thus, the glueball masses are not fixed at present. 
The glueball masses obtained here have been compared with the results obtained by the 
lattice QCD simulation. 
The reasonable results are included under a certain 
strength of QCD coupling constant. 
However, it should be necessary to 
investigate other glueball states such as $2^+$ state.
In addition to the glueballs with the other quantum numbers, it is interesting to study the excited glueball states. 
In order to calculate the excited glueball masses, the other trial states $\ket{\Phi'}$ may be 
introduced in which $\langle \Phi'\ket{\Phi}=0$ should be satisfied. 
This treatment is similar to that of the Hartree-Fock method to calculate the excited states in the nuclear many-body problem.  
Another possibility is to consider the three gluon states,\cite{Jaffe}
where the external source term consists of three gluons. 
These investigations rest future problems. 
Further, in this paper, the glueball masses at zero temperature were 
considered because $f_n$ is adopted as $1/2$ in Eq.(\ref{2-17}) or (\ref{2-19}).
However, if the eigenvalue of the reduced density matrix is calculated in the finite 
temperature $T$ in which $f_{\bf k}^{\sigma}=\sigma\cdot (n_{\bf k}^{\sigma}+1/2)$ is obtained 
where $n_{\bf k}^{\sigma}=1/(e^{\sigma E_{\bf k}/T}-1)$ is the bose distribution 
function, then, it may be possible to evaluate 
the glueball masses at finite temperature. 
These are future problems.

\section*{Acknowledgements} 
The author would like to express his sincere thanks to Professor 
K. Iida, Dr. E. Nakano, Dr. T. Saito and Dr. K. Ishiguro whose are 
the members of Many-Body Theory Group of Kochi University. 
The author also would like to express his sincere thanks to the late 
Professor Dominique Vautherin for the collaboration and giving him 
the suggestion for this work developed in this paper. 
He is partially supported by the Grants-in-Aid of the Scientific Research 
(No.23540311) from the Ministry of Education, Culture, Sports, Science and 
Technology in Japan.

\appendix
\section{Evaluation of the polarization tensor $\Pi_0(\omega,{\mib q})$ in the dimensional regularization scheme}

Let us show the polarization tensor in Eq.(\ref{4-1}) again:
\beq\label{a1}
\Pi_0(\omega,{\mib q})&\equiv& g^2{\wtilde \Pi}_0(\omega,{\mib q})\nonumber\\
&=&g^2\int\frac{d^3{\mib k}}{(2\pi)^3}\frac{1}{4E_{\bf k}E_{{\bf k}-{\bf q}}}
\left[\frac{1}{\omega-(E_{\bf k}+E_{{\bf k}-{\bf q}})}-\frac{1}{\omega+(E_{\bf k}+E_{{\bf k}-{\bf q}})}\right]\nonumber\\
&=&g^2\int\frac{d^4 k}{i(2\pi)^4}S_kS_{k-q}\ , \\
S_k&=&\frac{-i}{k^2+i\epsilon}\ .\nonumber
\eeq
Here, the 4-momentum integration is rewritten in the following form by 
using the Feynman parameter formula: 
\beq\label{a2}
\Pi_0(\omega,{\mib q})&=&-g^2\int\frac{d^4k}{i(2\pi)^4}\frac{1}{k^2+i\epsilon}\frac{1}{(k-q)^2+i\epsilon}\nonumber\\
&=&-g^2\int_0^1dx\int\frac{d^4k}{i(2\pi)^4}\frac{1}{\left[(k-q)^2x+k^2(1-x)\right]^2}\ .
\eeq
The integration of the right-hand side diverges. 
Thus, one needs to regularize the divergent integral to get the finite result. 
In this paper, the dimensional regularization method is applied and 
so-called modified minimal subtraction (${\ovl {\rm MS}}$) scheme is adopted 
with the consistency to the evaluation of the QCD running coupling constant $g_R^2$.\cite{Vautherin}   
Thus, the integration is regarded as the $n$-dimensional integration and 
is calculated as 
\beq\label{a3}
\Pi_0(\omega,{\mib q})&=&-g^2\int_0^1dx\int\frac{d^n k}{i(2\pi)^n}\frac{1}{\left[(k-q)^2x+k^2(1-x)\right]^2}\nonumber\\
&=&-\frac{g^2\Gamma(2-n/2)}{(4\pi)^{n/2}}\int_0^1dx\frac{1}{\left[q^2x(x-1)\right]^{2-n/2}}\nonumber\\
&=&-g^2(\mu^2)^{-(2-\frac{n}{2})}\cdot(-1)^{-(2-\frac{n}{2})}\cdot\frac{\Gamma(2-n/2)}{(4\pi)^{n/2}}\nonumber\\
& &\qquad\times
\int_0^1dx\frac{(\mu^2)^{2-n/2}}{(q^2)^{2-n/2}}\cdot
\frac{(-1)^{2-n/2}}{\left[x(x-1)\right]^{2-n/2}}\ , 
\nonumber\\
& &
\eeq
where $\Gamma(z)$ is the Gamma function. 
Here, we introduce a momentum scale $\mu$ to define the dimensionless coupling as follows. 
In $n$-dimension, QCD coupling $g$ has a dimension. 
Thus, the dimensionless coupling $g_R$ should be introduced. 
Then, we can further rewrite the above result as 
\beq\label{a4}
& &\Pi_0(q)=-g_R^2\frac{\Gamma(\epsilon)}{(4\pi)^2}(-1)^{-\epsilon}\left(\frac{\mu^2}{q^2}\right)^{\epsilon}
\int_0^1dx \left[\frac{4\pi}{x(1-x)}\right]^{\epsilon}\ \ , 
\eeq
where we define the dimensionless coupling\cite{Muta} as
\beq\label{a5}
& &g_R^2=g^2(\mu^2)^{-\epsilon}\ , \nonumber\\
& &\epsilon=2-\frac{n}{2} \ .
\eeq
Of course, if $n=4$, then $\epsilon=0$. 
Therefore, for infinitesimal value of $\epsilon$, we get 
\beq\label{a6}
\Gamma(\epsilon)&=&\frac{1}{\epsilon}-\gamma+O(\epsilon)\ , \nonumber\\
(4\pi)^{\epsilon}&=&1+\epsilon\ln 4\pi+O(\epsilon^2)\ , \nonumber\\
(-1)^{-\epsilon}&=&1+i\epsilon \pi +O(\epsilon^2)\ , \nonumber\\
\left(\frac{\mu^2}{q^2}\right)^{\epsilon}&=&1-\epsilon\ln \left(\frac{q^2}{\mu^2}\right)+O(\epsilon^2)\ , \nonumber\\
\int_0^1 dx \frac{1}{[x(1-x)]^\epsilon}&=&\int_0^1dx \left[1-\epsilon\ln[x(1-x)]\right]+O(\epsilon^2)\nonumber\\
&=&1+2\epsilon+O(\epsilon^2)\ .
\eeq
Thus, the polarization tensor $\Pi_0(q)\equiv g^2{\wtilde \Pi}_{0}(q)$ can be evaluated as 
\beq\label{a7}
\Pi_0(q)
&=&
\frac{g_R^2}{16\pi^2}\left[-\left(\frac{1}{\epsilon}-\gamma+\ln 4\pi\right)+\ln\frac{q^2}{e^2\mu^2}-i\pi+{O}(\epsilon)\right]\ . 
\eeq
Since we apply the ${\ovl {\rm MS}}$-scheme in order to get a finite value by subtracting the 
divergent term, 
we subtract a part proportional to the following set:
\beq\label{a8}
\frac{1}{\ovl \epsilon}\equiv \frac{1}{\epsilon}-\gamma+\ln 4\pi\ .
\eeq
Finally, as a result, we get the polarization tensor as 
\beq\label{a9} 
{\rm Re}\ \Pi_0(q)=\frac{g_R^2}{16\pi^2}\ln\frac{q^2}{e^2\mu^2}\ ,\qquad
{\rm Im}\ \Pi_0(q)=-\frac{g_R^2}{16\pi}\ . 
\eeq

\section{Decay width}

In this Appendix, the glueball mass is reconsidered by taking into account 
the imaginary part of the polarization tensor which may lead to the 
decay width of the glueball. 
In general, when a mass function $\Sigma(p)$ has an imaginary part, the propagator has a 
form 
\beq\label{c1}
S_{IJ}\propto \frac{1}{p^2-\Sigma^2}=\left[p^2-\left(M^2-\frac{\Gamma^2}{4}\right)+iM\Gamma\right]^{-1} \ ,
\eeq
where the mass function $\Sigma$ is divided into a real and an imaginary part as 
$\Sigma=M-i\Gamma/2$. 
Thus, the relations 
\beq\label{c2}
& &{\rm Re}\ S^{-1}(\omega, {\mib p}={\mib 0})=0\ , \nonumber\\
& &{\rm Im}\ S^{-1}(\omega,{\mib p}={\mib 0})=M\Gamma
\eeq
give the mass $M$ and $\Gamma$ that 
may be regarded as a decay width.

\begin{figure}[t]
\begin{center}
\includegraphics[height=5.1cm]{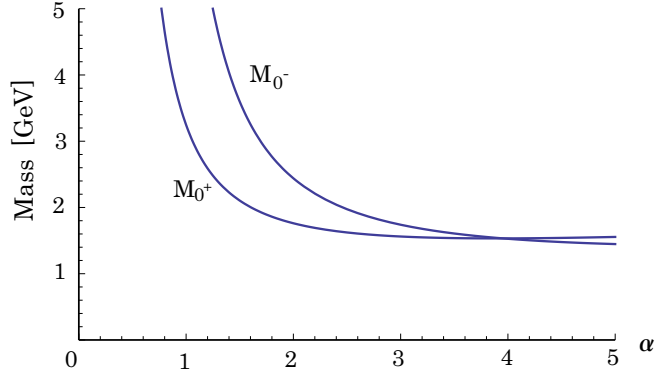}
\caption{
The scalar glueball mass $(M_{0^+})$ and the pseudoscalar glueball mass ($M_{0^-})$ 
are shown as a function of the QCD running coupling 
$\alpha_{\rm QCD}=g_R(\mu)^2/(4\pi)$, in which the imaginary part of the response function 
is taken into account. 
The QCD scale parameter is taken as $\Lambda_{\rm QCD}=0.20$ GeV.
}
\label{fig:4-2}
\end{center}
\end{figure}

In the treatment developed in this paper, the above relations are rewritten for the scalar $0^+$ 
and pseudoscalar $0^-$ glueballs as 
\beq\label{c3}
& &1-a_{\pm}c_3{\rm Re}\ {\Pi}_0(\omega_{0^{\pm}},{\mib 0})=0 \ , \qquad \left(\omega_{0^{\pm}}^2=M_{0^{\pm}}^2-\frac{\Gamma_{0^{\pm}}^2}{4}\right)
\nonumber\\
& &-a_{\pm}c_3{\rm Im}\ {\Pi}_0(\omega_{0^{\pm}},{\mib 0})=\frac{M_{0^{\pm}}\Gamma_{0^{\pm}}}{\omega_{0^{\pm}}^2}
\eeq
with $a_+=2$ for $0^+$ and $a_-=1$ for $0^-$ glueball. 
Here, 
\beq\label{c4} 
\Pi_0(q)\equiv g^2{\wtilde \Pi}_0(q)=\frac{g_R^2}{16\pi^2}\left(\ln\frac{q^2}{e^2\mu^2}-i\pi\right)\ .
\eeq
From Eq.(\ref{c2}), $M$ and $\Gamma$ are obtained as 
\beq
M_{0^{\pm}}&=&\frac{\mu e\cdot e^{\frac{8\pi^2}{3a_{\pm}g_R^2}}}
{\sqrt{1-\!\left(\frac{16\pi}{3a_{\pm}g_R^2}\right)^2\!\left(\!\!\sqrt{1+\left(\frac{3a_{\pm}g_R^2}{16\pi}\right)^2}-1\!\!\right)^2}}
\nonumber\\ 
&=&\frac{\Lambda_{\rm QCD}\cdot\exp\left(1+\frac{\pi}{\alpha_{\rm QCD}}\left(\frac{2}{3a_{\pm}}+\frac{1}{4\pi^2b_0}\right)\right)}
{\sqrt{1-\!\left(\frac{4}{3a_{\pm}\alpha_{\rm QCD}}\right)^2\!\left(\!\!\sqrt{1+\left(\frac{3a_{\pm}\alpha_{\rm QCD}}{4}\right)^2}-1\!\!\right)^2}}
\label{c5}\\
\Gamma_{0^{\pm}}&=&
\frac{8M_{0^{\pm}}}{3a_{\pm}\alpha_{\rm QCD}}\left(\sqrt{1+\left(\frac{3a_{\pm}\alpha_{\rm QCD}}{4}\right)^2}-1\right)\ . 
\label{c6}
\eeq
In Fig.3, the glueball masses are shown as a function of $\alpha_{\rm QCD}=g_R^2/4\pi$ with 
$\Lambda_{\rm QCD}=0.2$ GeV. 
If the QCD coupling constant $\alpha_{\rm QCD}$ is roughly taken as $\alpha_{\rm QCD}=2.0$, then the 
glueball masses are obtained as 
\beq\label{c7}
M_{0^+}=1.76\ {\rm GeV}\ , \qquad
M_{0^-}=2.44\ {\rm GeV}\ .
\eeq
Under the above parameter set, from the imaginary part of the polarization tensor, the 
decay width may be evaluated, which results $\Gamma_{0^+}=2.54$ GeV and $\Gamma_{0^-}=2.61$ GeV. 
These values are rather large, while a large decay width is reported by using a chiral quark model.\cite{decay} 
As is mentioned in \S 4, only gluons are contained in this framework. 
Then, it is only possible that the glueball decays to color-octet gluons, which should be forbidden due to 
the color confinement. 
However, in our treatment, the color confinement is not considered explicitly, 
so the glueballs easily decay to gluons. 
Thus, the rather large decay width, which means a decay from the color-singlet glueball to the 
color-octet gluons, may be obtained unavoidably. 
Thus, the investigation of the decay of glueball is still remained 
as an open question.

\section{Gluon mass}

In this appendix, we show that the gluon mass itself is zero in the framework developed in this paper 
under the Gaussian approximation used there. 
First, the canonical equations of motion for ${\ovl A}_i^a({\mib x},t)$ and ${\ovl E}_i^a({\mib x},t)$ in 
Eq.(\ref{2-9a}) with (\ref{2-7}) are given as 
\beq\label{b1}
{\dot {\ovl A}}_i^a(x)&=&{\ovl E}_i^a(x)\ , \nonumber\\
{\ddot {\ovl A}}_i^a(x)&=&{\dot {\ovl E}}_i^a(x)\nonumber\\
&=&-\epsilon_{ijk}\partial_j{\ovl B}_k^a(x)+g\epsilon_{ijk}f^{abc}{\ovl B}_j^b(x){\ovl A}_k^c(x)
-\frac{1}{2}\int d^3{\mib y}\frac{{\delta K}_{jl}^{bc}({\mib y})}{\delta{\ovl A}_i^a({\mib x},t)}G_{lj}^{cb}({\mib y},{\mib y}), \ \quad
\eeq
where we define 
\beq\label{b2}
\frac{{\delta K}_{jl}^{bc}({\mib y})}{\delta{\ovl A}_i^a({\mib x},t)}
&=&-g\epsilon_{ijm}\epsilon_{mlk}f^{abd}\delta^3({\mib x}-{\mib y})\left(\delta_{dc}\partial_k^{\bf y}-gf^{dce}{\ovl A}_k^e({\mib y},t)\right)
\nonumber\\
& &-g\epsilon_{lim}\epsilon_{mjk}f^{cad}\left(-\delta_{db}\partial_k^{\bf y}-gf^{dbe}{\ovl A}_k^e({\mib y},t)\right)\delta^3({\mib x}-{\mib y})
\nonumber\\
& &-g\epsilon_{jlm}\epsilon_{mik}f^{bcd}\left(-\delta_{da}\partial_k^{\bf y}-gf^{dae}{\ovl A}_k^e({\mib y},t)\right)\delta^3({\mib x}-{\mib y})\ .
\eeq
In order to get the gluon propagator, we have to introduce the external term $H_{\rm ext}$ in the Hamiltonian as 
\beq\label{b3}
H_{\rm ext}=\int d^3{\mib x}J_i^a({\mib x},t){A}_i^a({\mib x})
=J_i^a e^{-i\omega t}\int d^3{\mib x}e^{i{\bf q}\cdot{\mib x}}A_i^a({\mib x})\ .
\eeq
Thus, the gluon propagator $S_{ij}^{ab}$ can be derived by $\delta \langle A_i^a \rangle/\delta J_j^b$ following the 
general discussion. 

Let us start with solutions ${\ovl A}_i^a({\mib x},t)={\ovl E}_i^a({\mib x},t)=0$ under the Hamiltonian $H_0$. 
With the external term, the solutions should be shifted. 
Here, we denote them as 
\beq\label{b4}
{\ovl A}_i^a({\mib x},t)=0+\delta {\ovl A}_i^a({\mib x},t)\ , \qquad
{\dot {\ovl E}}_i^a({\mib x},t)=-\frac{\delta\langle H\rangle}{\delta {\ovl A}_i^a({\mib x})}=-J_i^a({\mib x})\ .
\eeq
From the equation of motion in (\ref{b1}), we can get the equation of motion for $\delta{\ovl A}_i^a({\mib x},t)$ 
with a linear approximation for $\delta{\ovl A}_i^a({\mib x},t)$ under small source current $J_i^a$: 
\beq\label{b5}
& &\left((\partial_t^2-{\mib \nabla}^2)\delta_{ij}+\partial_i\partial_j\right)\delta{\ovl A}_j^a({\mib x},t)
-\frac{c_3}{2}g^2(\delta_{il}\delta_{jk}-2\delta_{ik}\delta_{jl}+\delta_{ij}\delta_{kl})G_{lj}({\mib x})\delta{\ovl A}_k^a({\mib x},t)
\nonumber\\
&=&-J_i^a({\mib x})\ , 
\eeq
where we introduced a new notation $G_{lj}({\mib x})$ through $G_{lj}^{ab}({\mib x},{\mib x})\equiv \delta_{ab}G_{lj}({\mib x})$. 
In the above equation of motion, the second term, $G_{lj}({\mib x})$, is diagrammatically represented by so-called tadpole 
diagram. 
It is well known that there is no tadpole contribution in pure Yang-Mills gauge theory in the dimensional regularization scheme, namely, 
\beq\label{b6}
G_{lj}({\mib x})&\propto&
\int\frac{d^3{\mib k}}{(2\pi)^3}\frac{1}{2|{\mib k}|}=\int\frac{d^4 k}{i(2\pi)^4}\frac{-1}{k_0^2-|{\mib k}|^2+i\epsilon}
\nonumber\\
&\rightarrow&
\left.\int \frac{d^n k}{i(2\pi)^n}\frac{1}{m^2-k^2}\right|_{m^2\rightarrow 0}\nonumber\\
&=&\left.
\frac{1}{(4\pi)^{\frac{n}{2}}}\frac{\Gamma(\varepsilon)}{1-\frac{n}{2}}m^2(m^2-i\epsilon)^{-\varepsilon}
\right|_{m^2\rightarrow 0}\nonumber\\
&=&-\left.\frac{1}{(4\pi)^2}\left[\left(\frac{1}{\varepsilon}-\gamma\right)m^2-m^2\ln m^2+O(\varepsilon)\right]\right|_{m^2\rightarrow 0}
\nonumber\\
&=&0\ .
\eeq
Thus, from the equation of motion in Eq.(\ref{b5}), the following equation in the momentum space is obtained: 
\beq\label{b7}
& &\left[\omega^2\delta_{ij}-|{\mib q}|^2\left(\delta_{ij}-\frac{q_iq_j}{|{\mib q}|^2}\right)\right]\delta{\ovl A}_j^a=J_i^a\ .
\eeq
Finally, the gluon mass is given by the pole of the gluon propagator 
$S_{ij}^{ab}(\omega,{\mib q})={\delta{\ovl A}_i^a}/{J_j^b}$, 
namely, gluon mass is exactly zero in this framework.

\end{document}